\documentclass{article}

\usepackage{arxiv}

\usepackage[utf8]{inputenc} 
\usepackage[T1]{fontenc}    
\usepackage{hyperref}       
\usepackage{url}            
\usepackage{booktabs}       
\usepackage{amsfonts}       
\usepackage{nicefrac}       
\usepackage{microtype}      
\usepackage{lipsum}
\usepackage{graphicx}
\graphicspath{ {./images/} }
\usepackage{amsmath}
\usepackage{amssymb}
\usepackage{amsthm}

\usepackage{booktabs}
\usepackage{verbatim}
\usepackage{xcolor}
\usepackage{amstext}
\usepackage{booktabs}
\usepackage{upgreek}
\usepackage{multirow}
\usepackage{bbm}
\usepackage[algo2e,ruled,vlined]{algorithm2e}

\newcommand{\beq}{\begin{equation}}
\newcommand{\eeq}{\end{equation}}

\renewcommand{\vec}[1]{\mathbf{#1}}
\newcommand{\mr}[1]{\mathrm{#1}}

\newcommand{\expect}{\mathbb{E}}
\newcommand{\real}{\mathbb{R}}
\newcommand{\ttimes}{\,$\times$\,}

\newcommand{\loss}{\mathcal{L}}

\newcommand{\bb}{\vec{b}}
\newcommand{\ff}{\vec{f}}
\newcommand{\xx}{\vec{x}}
\newcommand{\yy}{\vec{y}}
\newcommand{\kk}{\vec{k}}
\newcommand{\dd}{\vec{d}}

\newcommand{\finv}{f^{\mr{inv}}}

\newcommand{\myparagraph}[1]{\noindent\textbf{#1}~~}

\newcommand{\fig}[1]{\includegraphics[width=0.19\linewidth]{figure/#1}} 

\title{Privacy Preserving for Medical Image Analysis\\ via Non-Linear Deformation Proxy}

\author{
 Bach Ngoc Kim \\
  Software Engineering and IT Department\\
  ETS Montreal\\
  Canada H3C 1K3 \\
  \texttt{bachknk49@gmail.com} \\
   \And
 Jose Dolz \\
  Software Engineering and IT Department\\
  ETS Montreal\\
  Canada H3C 1K3 \\
  \texttt{jose.dolz@etsmtl.ca} \\
  \And
  Christian Desrosiers \\
  Software Engineering and IT Department\\
  ETS Montreal\\
  Canada H3C 1K3 \\
  \texttt{christian.desrosiers@etsmtl.ca} \\
  \And
  Pierre-Marc Jodoin \\
 Computer Science Department\\
  Sherbrooke University\\
  Canada J1K 2R1 \\
  \texttt{pierre-marc.jodoin@usherbrooke.ca} \\
}

\begin{document}
\maketitle
\begin{abstract}
We propose a client-server system which allows for the analysis of multi-centric medical images while preserving patient identity. In our approach, the client protects the patient identity by applying a pseudo-random non-linear deformation to the input image. This results into a proxy image which is sent to the server for processing. The server then returns back the deformed processed image which the client reverts to a canonical form. Our system has three components: 1) a flow-field generator which produces a pseudo-random deformation function, 2) a Siamese discriminator that learns the patient identity from the processed image, 3) a medical image processing network that analyzes the content of the proxy images. The system is trained end-to-end in an adversarial manner. By fooling the discriminator, the flow-field generator learns to produce a bi-directional non-linear deformation which allows to remove and recover the identity of the subject from both the input image and output result. After end-to-end training, the flow-field generator is deployed on the client side and the segmentation network is deployed on the server side. The proposed method is validated on the task of MRI brain segmentation using images from two different datasets. Results show that the segmentation accuracy of our method is similar to a system trained on non-encoded images, while considerably reducing the ability to recover subject identity.
\end{abstract}


\begin{figure*}[tp]
 \centering
 \includegraphics[width=.85\textwidth]{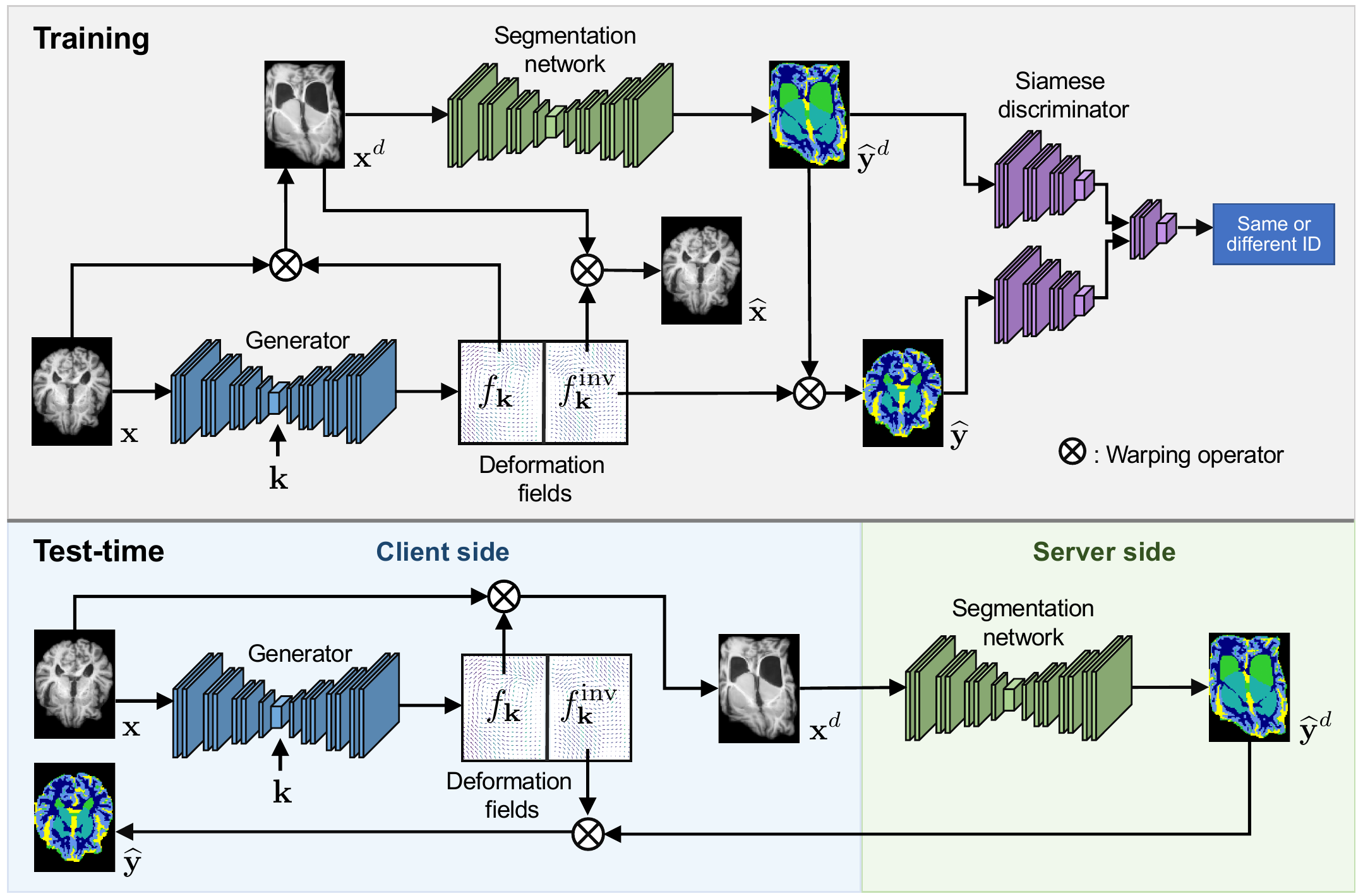} \\
 \caption{System diagram for training (\textit{top}) and testing (\textit{bottom}).}
 \label{fig:system_training}
\end{figure*}


\section{Introduction}


Convolutional neural networks (CNNs) are the {\em de facto} solutions to a large number of medical image analysis tasks, from disease recognition, to anomaly detection, segmentation, tumor resurgence prediction, and many more~\cite{Wang2016,dolz20183d,Lee2020,litjens2017survey}. While solutions to these decade long problems are flourishing, a consistent obstacle to their deployment has been privacy protection.  
Despite being essential to preserve civil rights, privacy protection rules are nonetheless a break on the development of machine learning methods, and in particular to cloud-based medical image analysis solutions. However, cloud-based solutions have great benefits, such as preventing clinics from having to purchase and maintain specialized hardware. As such, if these systems are to prosper in the medical world, they will have to integrate privacy protection policies to their processes.  

The simplest privacy protection protocol is anonymization. For medical images, this means removing patient tags from DICOM images or converting it into identity-agnostic formats such as TIFF. Unfortunately, patient identity can be recovered just by inspecting raw images~\cite{Kumar209726,kim2020privacynet}. Results reported in Section~\ref{sec:experiment} show that the identity-recognition F1-scores can go up to 98\%. Needless to say, data exchanged between the client and the server can be encrypted. While this ensures protection against outside cybercriminals, it does not protect against malicious people from within the organization. Alternatively, one can use homomorphic encryption which allows to perform forward and backward passes of encrypted data without having to decrypt it~\cite{ziad2016cryptoimg,nandakumar2019towards}. Although these methods perform well in some applications, homomorphic cryptosystems typically incur high computational costs \cite{Hesamifard17,nandakumar2019towards,paillier1999public} and are mostly restricted to simple neural networks.



In this paper, we propose a novel client-server system which can process medical images while preserving patient identity. As shown in Fig.~\ref{fig:system_training}, instead of sending an image $\xx$ to the server, the client deforms the image with a non-linear spatial deformation field $f_{\kk}$ conditioned on a client-specific private key $\kk$. The warped image $\xx^d$ is then sent to the server where it is processed and sent back to the client. At the end, the deformed result $\yy^d$ is unwarped with the inverse transformation function $\finv$. Results obtained on the task of 3D MRI brain image segmentation reveal that the patient identity is preserved both on the MRI image and the segmentation map while keeping a high segmentation accuracy.


\section{Related works}


\myparagraph{Homomorphic encryptions} 
One way of preserving privacy is via homomorphic-encryption  (HE)~\cite{Dowlin16,Hesamifard17,nandakumar2019towards}, which allows neural networks to process encrypted data without having to decrypt it. However, HE is not void of limitations. First, it has non-negligible communication overhead~\cite{rouhani2018}.  
Furthermore, being limited to multiplications and additions, non-linear activation functions have to be approximated by polynomial functions, which makes CNNs prohibitively slow (Nandakumar {\em et al.}~\cite{nandakumar2019towards} report processing rates of 30 min per image). Thus, homomorphic networks have been relatively simplistic~\cite{Hardy2017} and it is not clear how state-of-the-art deep neural nets~\cite{DBLP:journals/corr/RonnebergerFB15} can accommodate this approach.  


\myparagraph{Federated learning} Another solution for multi-centric deep learning data analysis is federated learning. \cite{DBLP:journals/corr/XieBFGLN14,DBLP:journals/corr/KonecnyMRR16,DBLP:journals/corr/McMahanMRA16,DBLP:journals/corr/abs-1812-03288,yang2019federated, McMahan2017}. The idea of this approach is to train a centralized model by keeping the data of different clients decentralized and exchanging model parameters or backpropagated gradients during training. While it improves privacy by not sharing data, this approach requires significant network bandwidth, memory and computational power, and is susceptible to data leakage from specialized attacks like model inversion~\cite{wu2019p3sgd,zhu2019deep}. Most of all, in such approach, private data must still be sent at test time from clients to the server, which defeats the purpose.


\myparagraph{Image deformation in medical imaging} Spatial deformable transformations have been widely employed in the medical field, mainly for the registration task \cite{sotiras2013deformable}. While initial attempts were based on non-learning methods, e.g., elastic models \cite{shen2002hammer} or B-splines \cite{klein2007evaluation}, recent works leverage deep neural networks to learn a function for the registration \cite{DBLP:journals/corr/abs-1809-05231}. More recently, spatial deformations have been used in the context of medical image segmentation to augment the training data by synthesizing new training examples \cite{DBLP:journals/corr/abs-1902-05396,Zhao_2019_CVPR}. In contrast, we resort to these models with the goal of distorting input volumes, so that the structural information required to recover the user identity cannot be identified in the server side.  


\myparagraph{Privacy preserving with adversarial learning}
A popular solution consists in training a generator to create perturbed images, from either a noise distribution \cite{xu2019ganobfuscator} or real images \cite{raval2017protecting}. Then, the generated images are employed to train a discriminator to differentiate between original and synthetic images. Nevertheless,  the encoding in these frameworks is not optimized under the supervision of specific utility objectives, potentially achieving sub-optimal results and sacrificing the performance on the utility task. To overcome this limitation, recent works have integrated specific utility losses, which are jointly optimized with the privacy objectives \cite{pittaluga2019learning,wu2018towards,yang2018learning,DBLP:journals/corr/abs-1904-05514,ren2018learning,xiao2019adversarial}. These approaches, which typically tackle simple problems (i.e., QR code classification or face recognition), resort to standard classification models for both the adversarial and task-specific objectives, where the number of classes is fixed. 
An alternative to alleviate the issue when the number of classes is non-fixed is to employ a Siamese architecture as the discriminator, which predicts whether two encoded images come from the same subject \cite{oleszkiewicz2018siamese,kim2020privacynet}. 


\myparagraph{Differences with existing methods} In contrast with prior works, the proposed framework can easily scale-up to non-fixed classes scenarios. Furthermore, compared to \cite{oleszkiewicz2018siamese}, our approach presents significant differences both in the objectives and methodology. First, privacy preserving is investigated in the context of biometrical data in \cite{oleszkiewicz2018siamese} (e.g., fingerprint), whereas we focus on volumetric medical images, which are dissimilar in nature. Second, they aim at finding the smallest possible transformation of an image to remove identity information while can be still used by non-specific applications. In contrast, our goal is to obfuscate images with the strongest possible transformation so that subject identity cannot be recovered while at the same time the encoded image can be used to train a model in the segmentation task. This results in important methodological differences, such as an additional network and different objective functions. More related is the work in \cite{kim2020privacynet} 
whose transformations in deteriorated images come in the form of intensity changes. But contrary to our method, the structural information in the segmentation results is preserved, which can be used to retrieve the patient identity. 

\begin{figure}[tp]
 \centering
 \includegraphics[width=.7\linewidth]{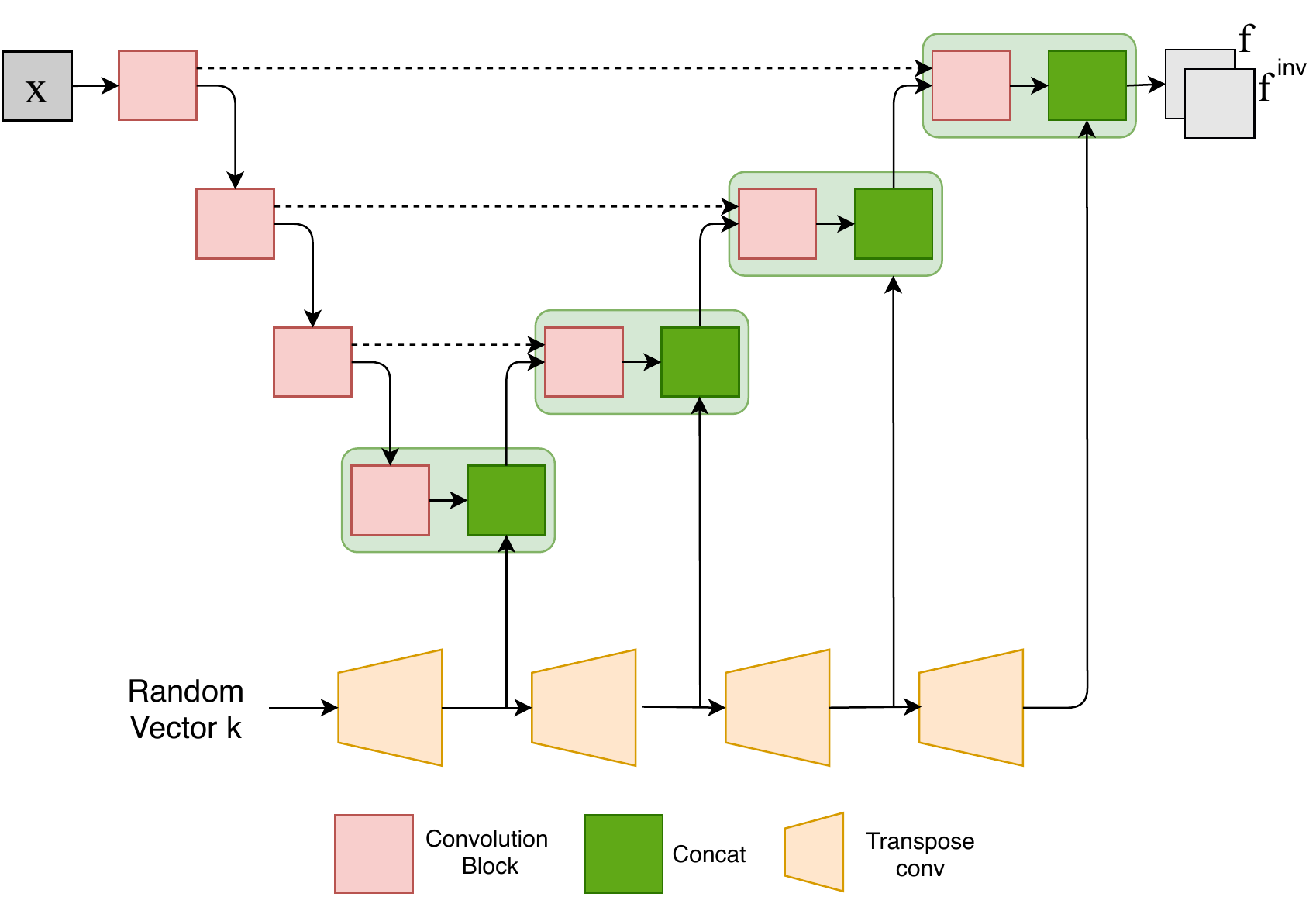}
 \caption{Network architecture of the generator.}
 \label{fig:architecture}
\end{figure}

\section{Methods}

We start by presenting the proposed architecture for privacy-preserving segmentation. We then describe the transformation to distort images and the loss functions used for training our model.

\subsection{Proposed architecture}
\label{sec:architecture}

As shown in Fig.~\ref{fig:system_training}, during training, our system consists of three components: a {\em transformation generator}, a {\em segmentation network} and a {\em discriminator}. We describe the role of each of these components below. 

\subsubsection{Transformation generator} 

The first component is a generator $G$ that takes as input a 3D image $\xx \in \real^{H\times W\times D}$ and a random vector $\kk \in \real^{M}$ and outputs a transformation $f_{\kk}$ that distorts $\xx$ so that the corresponding subject's identity cannot be recovered, yet segmentation can still be performed. Vector $\kk$ is a private key, known only by the client, that parameterizes the transformation function and ensures that this function cannot be inferred only from distorted images.   

In Privacy-Net~\cite{kim2020privacynet} a generator is also used for this purpose, however, it directly outputs the distorted image. In this work, we follow a different approach where the generator outputs the transformation function $f_{\kk}$, which is used afterwards to distort the image. Computing this function explicitly enables to perform the segmentation in the \emph{transformed space}, where identity is obfuscated, and then reverse the transformation back to the original space. To ensure that the transformation is reversible, we could limit $f_{\kk}$ to a specific family of functions (e.g., free-form deformation~\cite{Wolberg99}). However, to add flexibility and learn a function most suitable for the downstream segmentation, we instead enforce the generator to output both $f_{\kk}$ and its inverse $\finv_{\kk}$, and use a reconstruction loss (see Section \ref{sec:invertibility_loss}) to impose that $\finv_{\kk}\!\circ\!f_{\kk} = I$. Given a training example $(\xx,\yy)$, where $\yy \in \real^{H\times W\times D \times C}$ is the ground-truth segmentation mask over $C$ classes, $f_{\kk}$ is used to compute the distorted image $\xx^d = f_{\kk}(\xx)$ and distorted segmentation $\yy^d = f_{\kk}(\yy)$. The former is sent to the segmentation network for processing, while $\yy^d$ is used to evaluate the segmentation output. On the other hand, the inverse function $\finv_{\kk}$ is used to obtain the reconstructed image $\widehat{\xx} = \finv_{\kk}(\xx^d)$ and reconstructed segmentation $\widehat{\yy} = \finv_{\kk}(\widehat{\yy}^d)$ in the original space.

The generator's architecture is shown in Fig.~\ref{fig:architecture}. It comprises an encoder path with 4 convolution blocks that takes an input image and computes feature maps of increasingly-reduced dimensions via pooling operations, and a decoder path also with 4 convolution blocks which produces an output map of same size as the input. In this work, we use the generator to predict a flow-field $f$ which assigns a displacement vector $f_{u,v,w} \in \real^3$ to each voxel $(u,v,w)$ of the 3D image $\xx$. More information on the transformation is given in Section \ref{sec:transformation_function}. Transpose convolutions are used in the decoder path to upscale feature maps. We also preserve high-resolution information by adding skip connections between convolution blocks at the same level of the encoder and decoder paths. Moreover, to ensure that the private key $\kk$ is used at different scales, we include another path in the model that gradually upscales $\kk$ with transpose convolutions and concatenates the resulting map with feature maps of corresponding resolution in the decoder path.     
%

\subsubsection{Segmentation network} 

The segmentation network $S$ takes as input the distorted image $\xx^d$ and outputs a distorted segmentation map $\widehat{\yy}^d = S(\xx^d)$. Although any suitable network can be employed, we used a 3D U-Net~\cite{cciccek20163d} which implements a convolutional encoder-decoder architecture with skip-connections between corresponding levels of the encoder and decoder. 


\subsubsection{Siamese discriminator} 

An adversarial approach is employed to obfuscate the identity of subjects in distorted images and segmentation maps.  In a standard approach, a classifier network is used as discriminator $D$ to predict the class (i.e., subject ID) of the encoded image produced by the generator. In our context, where the number of subjects can be in the thousands and grows over time, this approach is not suitable.  Alternatively, we follow a strategy similar to Privacy-Net~\cite{kim2020privacynet} where we instead use a Siamese discriminator that takes as input two segmentation maps, $\yy_i$ and $\yy_j$, and predicts whether they belong to the same subject or not. Note that this differs from Privacy-Net, which applies the Siamese discriminator on the encoded images, not on the segmentation maps. For training, we generate pairwise labels $s_{ij}$ such that $s_{ij}=1$ if $\yy_i$ and $\yy_j$ are from the same subject, otherwise $s_{ij}=0$. Since we now solve a binary prediction task, which is independent of the number of subjects IDs, this strategy can scale to a large and increasing number of subjects in the system. 


\subsubsection{Test-time system}
 
At testing, the system can be used for privacy-preserving segmentation as illustrated in Fig.~\ref{fig:system_training}. A client-side generator is first used with the client's private key $\kk$ to distort the 3D image to segment, $\xx$, into an identity-obfuscated image $\xx^d\!=\!f_{\kk}(\xx)$, which is then sent to the server for segmentation. The server-side segmentation network takes $\xx^d$ as input and outputs the distorted segmentation map $\widehat{\yy}^d$. Finally, $\widehat{\yy}^d$
is sent back to the client where the inverse transform is used to recover the segmentation map $\widehat{\yy} = \finv_{\kk}(\widehat{\yy}^d)$. 

\subsection{Transformation function}
\label{sec:transformation_function}

As in \cite{DBLP:journals/corr/abs-1809-05231,DBLP:journals/corr/abs-1902-05396}, the transformation function in our model takes an image (or segmentation map) and a flow-field as input, and outputs the deformed version of the image. Similarly to the spatial transformer network~\cite{DBLP:journals/corr/JaderbergSZK15}, this geometric deformation is based on grid sampling. Let $\bb$ be the base-grid of size $(H,W,D,3)$ containing the coordinates of image voxels, and $\ff$ be the deformation flow-field of same size. The coordinates of the deformed grid are then given by $\dd=\bb+\ff$. We obtain the deformed image $\xx^d$ by sampling the 8 neighbor voxels around each point of $\dd$ using tri-linear interpolation:
\begin{align}\label{eq:spatial-transf}
& x^d_{u,v,w} \, = \, \!\!\sum_{(u',v',w')\in \Omega} \!\!\!\!\!x_{u',v',w'}\cdot \max\big(0, 1\!-\!|d_u\!-\!b_{u'}|\big)\cdot\max\big(0, 1\!-\!|d_v\!-\!b_{v'}|\big)\cdot\max\big(0, 1\!-\!|d_w\!-\!b_{w'}|\big).
\end{align}
Since the function in Eq. (\ref{eq:spatial-transf}) is differentiable, we can back-propagate gradients during optimization.

\subsection{Training the proposed model}

We train the transformation generator $G$, segmentation network $S$ and discriminator jointly with the following five-term loss function :
\begin{equation}\label{eq:total_loss}
\begin{aligned}
& \loss_{\mr{total}}(S,G,D) \ = \ \loss_{\mr{seg}}(S) \, + \, \lambda_1 \loss_{\mr{adv}}(G,D) \, + \, \lambda_2 \loss_{\mr{inv}}(G) \, + \, \lambda_3 \loss_{\mr{smt}}(G) \, + \, \lambda_4 \loss_{\mr{div}}(G)
\end{aligned}
\end{equation}
Where $\lambda_1$, $\lambda_2$, $\lambda_3$ and $\lambda_4$ are hyper-parameters balancing the contribution of each term. In the following subsections, we define and explain the role of each term in this loss function.

\subsubsection{Segmentation loss}
\label{sec:segmentation_loss}

The segmentation loss enforces that the segmentation network $S$ learns a correct mapping from a distorted image $\xx^d\!=\!f_{\kk}(\xx)$ to its distorted segmentation $\widehat{\yy}^d$. The  predicted segmentation after reconstruction is $\widehat{\yy}=\finv_{\kk}(\widehat{\yy}^d)$.
Here, we use Dice loss~\cite{sudre2017generalised} to measure the difference between the reconstructed predicted segmentation and its corresponding ground-truth:
\begin{align}
\label{eq:segmentation_loss}
 \loss_{\mr{seg}}(S) & \, = \, \min_{S} \ \expect_{(\xx,\yy),\kk}\big[\ell_{\mr{\textsc{d}ice}}(\yy,\widehat{\yy})\big] \\
 & \, = \, \min_{S}\ \expect_{(\xx,\yy),\kk}\big[\ell_{\mr{\textsc{d}ice}}\big(\yy,(\finv_{\kk}\!\circ\!S\!\circ\!f_{\kk})(\xx)\big)\big].\nonumber
\end{align}
Since this loss samples over both images $\xx$ and random key vectors $\kk$, the network $S$ learns a segmentation that accounts for the variability of structures in images and their possible deformation resulting from $f_{\kk}$.  
\subsubsection{Identity obfuscation loss}
\label{sec:obfuscation_loss}

An adversarial loss is added to ensure that the transformation obfuscates subject identity.  
By maximizing the discriminator's error, the generator learns to produce transformed images from which identity cannot be recovered. However, this strategy is sensitive to noise or variation in contrast which ``fools'' the discriminator but still preserves structural information that can identify subjects. To alleviate this problem, we instead apply the discriminator on pairs of segmentation maps. Letting $D(\yy_i, \yy_j)$ be the probability that $\yy_i$ and $\yy_j$ are from the same subject, we define this loss as
\begin{align}\label{eq:loss_adv}
 & \loss_{\mr{adv}}(G,D) \, = \, \min_{G} \, \max_{D} \ \expect_{\yy_i,\yy_j}\big[s_{ij}\log D(\yy_i,\yy_j) \, + \, (1\!-\!s_{ij})\log\big(1- D(\yy_i,\yy_j)\big)\big] \, + \, \expect_{\yy,\kk}\big[\log\big(1 - D(\widehat{\yy}^d,\widehat{\yy})\big)\big]
\end{align}
with $\widehat{\yy}^d\!=\!S(f_{\kk}(\xx))$ and $\widehat{\yy}\!=\!\finv_{\kk}(\widehat{\yy}^d)$.  The first term corresponds to the cross-entropy loss on ground-truth segmentation pairs, that does not depend on the generator or segmentation network. The second  term measures the discriminator's ability to recognize that a deformed segmentation and its reconstructed version (by applying the reverse transform function) are from the same subject. This second term is optimized adversarially for $G$ and $D$. It can be shown using a variational bound method that optimizing the problem in Eq.~(\ref{eq:loss_adv}) minimizes the mutual information between a pair $(\widehat{\yy}^d, \widehat{\yy})$ and the same-identity variable $s_{ij}$~\cite{kim2020privacynet}. Consequently, it impedes a potential attacker from retrieving subject identity for a given distorted image by matching it with a database of existing images. 

\subsubsection{Transformation invertibility loss}\label{sec:invertibility_loss}

When receiving the deformed segmentation from the server, the client needs to bring it back to the original image space. For this to be possible, the transformation function needs to be invertible, i.e. $\finv\!\circ\!f = I$. To have this property, we minimize the $\loss_{\mr{\textsc{d}ice}}$ between a segmentation map and its reconstructed version. However, since the segmentation map is binary, this leads to non-smooth gradients. We avoid this problem by also minimizing the reconstruction error for input images, based on the structural similarity (SSIM) measure:
\begin{align}
\label{eq:invert_loss}
 \loss_{\mr{inv}}(G) \, = \, & \min_{G} \ \expect_{(\xx,\yy),k}\Big[\ell_{\mr{\textsc{ssim}}}\big(\xx,(\finv_{\kk}\!\circ\!f_{\kk})(\xx)\big) \, + \, \ell_{\mr{\textsc{d}ice}}\big(\yy, (\finv_{\kk}\!\circ\!f_{\kk})(\yy)\big)\big].
\end{align} 
where  $\ell_{\mr{\textsc{ssim}}}(\xx,\yy) \in [0,1]$ is the SSIM loss as  in~\cite{Zhao2017}.

The global SSIM is generated at each voxel using a 11$\times$11$\times$11 window, and then taking the average over all voxels. In practice, we use a multi-scale structural similarity (MS-SSIM) which computes the SSIM at multiple image scales via subsampling~\cite{wang2003multiscale}. 

\begin{table*}[tp]
\centering
\caption{Segmentation and re-identification results on the PPMI dataset.}
\label{table:Results}
\small
\setlength{\tabcolsep}{6pt}
\begin{tabular}{lcccccccccc}
\toprule
 & \multicolumn{6}{c}{Segmentation DSC} & \multicolumn{2}{c}{Re-id. F1-score} & \multicolumn{2}{c}{Re-id. mAP} \\
\cmidrule(l{5pt}r{5pt}){2-7}\cmidrule(l{5pt}r{5pt}){8-9}\cmidrule(l{5pt}r{5pt}){10-11}
Method & Overall & GM & WM & Nuclei & int.CSF & ext.CSF & Image & Seg. & Image & Seg. \\
\midrule
No-Proxy & 0.887 & 0.941 & 0.862 & 0.727 & 0.745 & 0.825 & 0.988 & 0.986 & 0.998 & 0.998 \\
Privacy-Net~\cite{kim2020privacynet} & 0.812 & 0.925 & 0.824 & 0.580 & 0.598 & 0.752 & -- & -- & 0.189 &0.632\\
\midrule
Ours (All losses) & 0.816 & 0.901 & 0.829 & 0.634 & 0.651 & 0.735 & 0.051 & 0.045 & 0.096 & 0.091 \\
Ours (w/o Invertibily) & 0.511 & 0.523 & 0.507 & 0.467 & 0.423 & 0.534 & 0.038 & 0.025 & 0.059 & 0.034 \\
Ours (w/o Smoothness) & 0.701 & 0.801 & 0.706 & 0.455 & 0.431 & 0.605 & 0.059 & 0.043 & 0.110 & 0.088 \\
Ours (w/o Diversity) & 0.864 & 0.934 & 0.853 & 0.718 & 0.711 & 0.796 & 0.445 & 0.473 & 0.393 & 0.329 \\
\bottomrule
\end{tabular} 
\end{table*}

\subsubsection{Transformation smoothness loss}
\label{sec:smoothness_loss}

The transformation invertibility loss in Eq. (\ref{eq:invert_loss}) may sometimes lead to discontinuity in the deformation field which prevents the segmentation from being reconstructed. To regularize the deformation field produced by the generator, we include another loss that enforces spatial smoothness:
\begin{align}
\label{eq:smoothness}
 \loss_{\mr{smt}}(G) \, = \, \expect_{\xx,\kk} \bigg[ \frac{1}{|\Omega|}\sum_{(u,v,w)\in\Omega}\!\!\!\! \| \nabla f_{u,v,w} \|_2\bigg].
\end{align}
The spatial gradient $\nabla f_{u,v,w}$ at each voxel $(u,v,w)$ is estimated using finite difference. 

\subsubsection{Transformation diversity loss}
\label{sec:diversity_loss}

A final loss in our model is added to prevent mode-collapse in the generator where the same transformation would be generated regardless of the input private key $\kk$. As mentioned before, having a transformation that depends on $\kk$ is necessary to avoid an attacker learn to ``reverse'' the transformation by observing several deformed images or segmentation maps. To achieve this, we maximize the distortion between two deformed versions of the same image or segmentation, generated from different random private keys $\kk$ and $\kk'$:
\begin{align}
\label{eq:distorion_loss}
 \loss_{\mr{div}}(G) & \, = \, \max_{G} \ \expect_{(\xx,\yy),\kk,\kk'}\big[\ell_{\mr{\textsc{ssim}}}(f_{\kk}(\xx),f_{\kk'}(\xx))\, + \, \ell_{\mr{\textsc{d}ice}}\big(f_{\kk}(\yy),f_{\kk'}(\yy)\big)\big].
\end{align}


\section{Experimental setup}
 
\subsection{Data set}

We evaluate our method on the task of privacy-preserving brain MRI segmentation. Two datasets are used in our experiments: the Parkinson's Progression Marker Initiative (PPMI) dataset~\cite{marek2011ppmi} and MRBrainS13 Challenge ~\cite{mendrik2015mrbrains} dataset. The first dataset, which contains longitudinal data, was considered for training the Siamese discriminator to recognize same-subject brain segmentations. The second one is used to evaluate the ability of our generator trained on PPMI to generalize to another dataset. 

\myparagraph{PPMI}This dataset contains T1 3D MRI scans from 350 subjects, acquired on 3T Siemens scanners from 32 different clinics. Each subject has 1--2 baseline acquisitions and 1--2 follow-up acquisitions a year later, resulting in a total of 773 images. We registered the MRI onto a common MNI space and reshaped it to a volume of 72\ttimes96\ttimes80 voxels with a 2\ttimes2\ttimes2\,mm$^3$ resolution. We split the data based on subjects, using 75\% of subjects (269 subjects, 592 images) for the training set and remaining 25\% (81 subjects, 181 images) for the testing set. As in~\cite{kim2020privacynet}, for the segmentation task, we used the labels generated automatically by Freesurfer for five classes: internal cerebrospinal fluid (CSF int), external cerebrospinal fluid (CSF ext), white matter (WM), gray matter (GM), and nuclei. To set the hyperparameters of our system, we followed a 5 fold cross-validation strategy on the training images. Once these hyperparameters were selected, we retrained the system on the entire training set and report results from the test set.

\myparagraph{MRBrainS} This second dataset comprises T1 3D MRI scans of 5 subjects obtained with a 3T Philips Achieva scanner, along with ground-truth segmentation masks for three classes: WM, GM and CSF. Note that the MRBrainS\,2013 Challenge included data for 15 additional subjects, however ground-truth segmentation was not provided for these subjects. As for PPMI, the MRBrainS images were registered to the MNI space using ANTs~\cite{Avants2011}. Since it has only 5 images, we directly report the cross-validation performance for this dataset.  

\begin{table*}[tp]
\centering
\caption{Influence of the different terms of the loss function ${\cal{L}}_{\mr{total}}$ 
in the reconstruction.}
\label{table:Reconstruction_Results}
\begin{small}
\begin{tabular}{lccccccc}
\toprule
 &  & \multicolumn{6}{c}{Segmentation DSC} \\
\cmidrule(l{5pt}r{5pt}){3-8}
 Method & MS-SSIM & Overall & GM & WM & Nuclei & int.CSF & ext.CSF \\
\midrule
Ours (All losses) & 0.993 & 0.983 & 0.987 & 0.982 & 0.970 & 0.972 & 0.985\\
Ours (w/o Invertibility) & 0.692 & 0.574 & 0.581 & 0.579 & 0.569 & 0.565 & 0.584\\
Ours (w/o Smoothness) & 0.905 & 0.829 & 0.856 & 0.861 & 0.822 & 0.791 & 0.842\\
Ours (w/o Diversity Loss) & 0.995 & 0.990 & 0.994 & 0.989 & 0.981 & 0.980 & 0.990\\
\bottomrule
\end{tabular} 
\end{small}
\end{table*}

\begin{figure*}[tp]
\begin{center}
\begin{footnotesize}
\begin{tabular}{cccc}
 \includegraphics[width=0.225\linewidth]{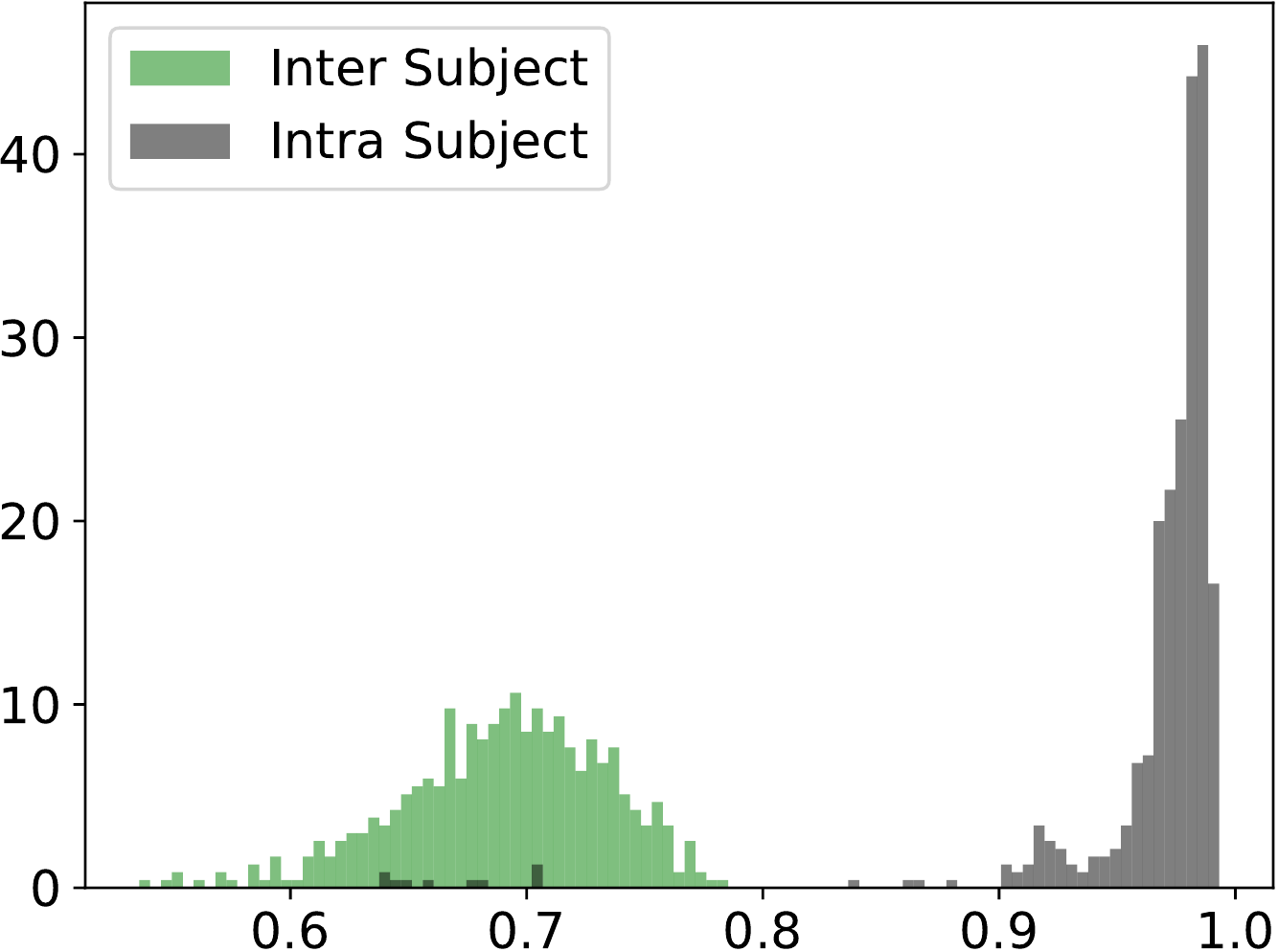} & \includegraphics[width=0.225\linewidth]{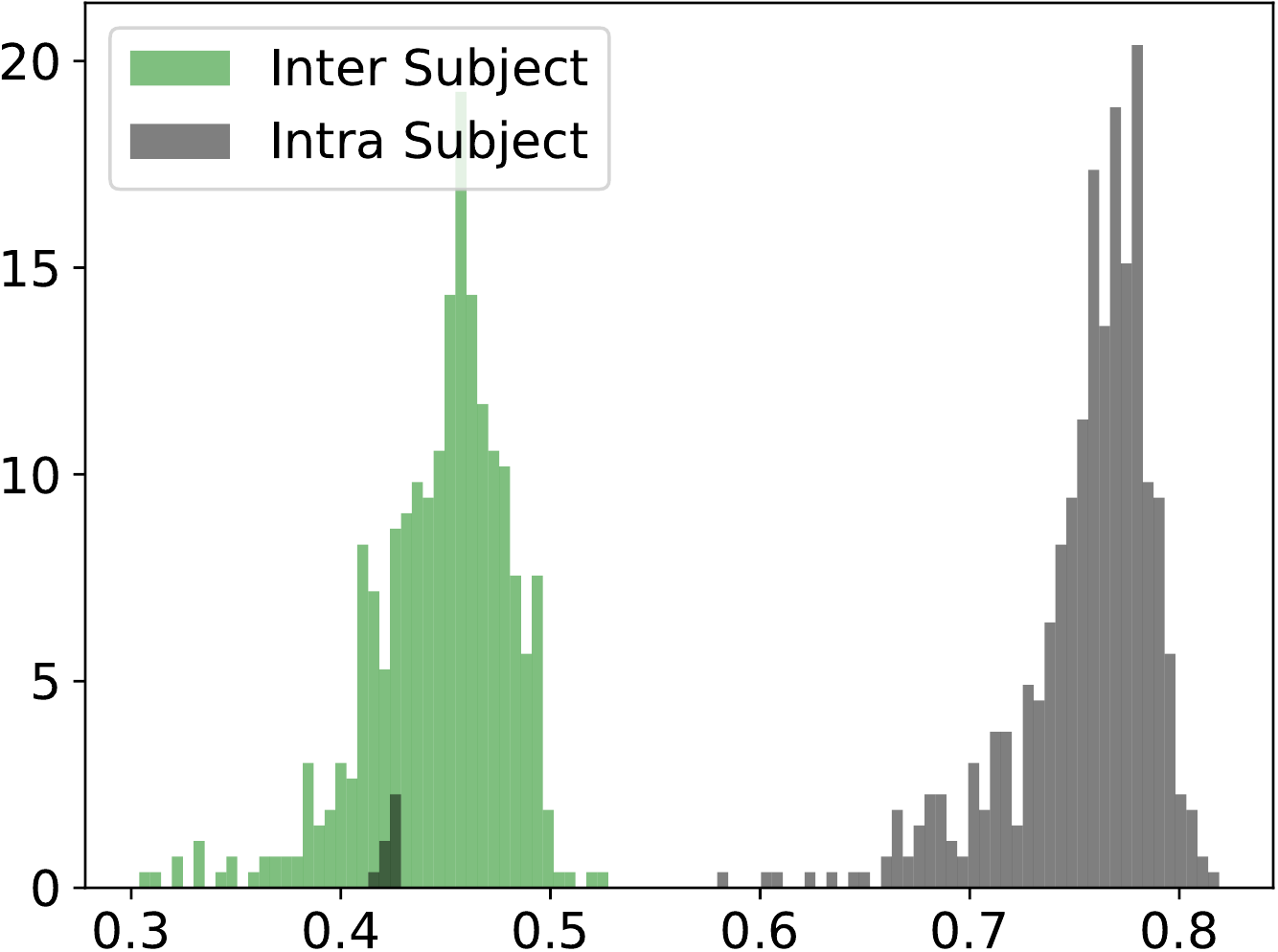} & \includegraphics[width=0.225\linewidth]{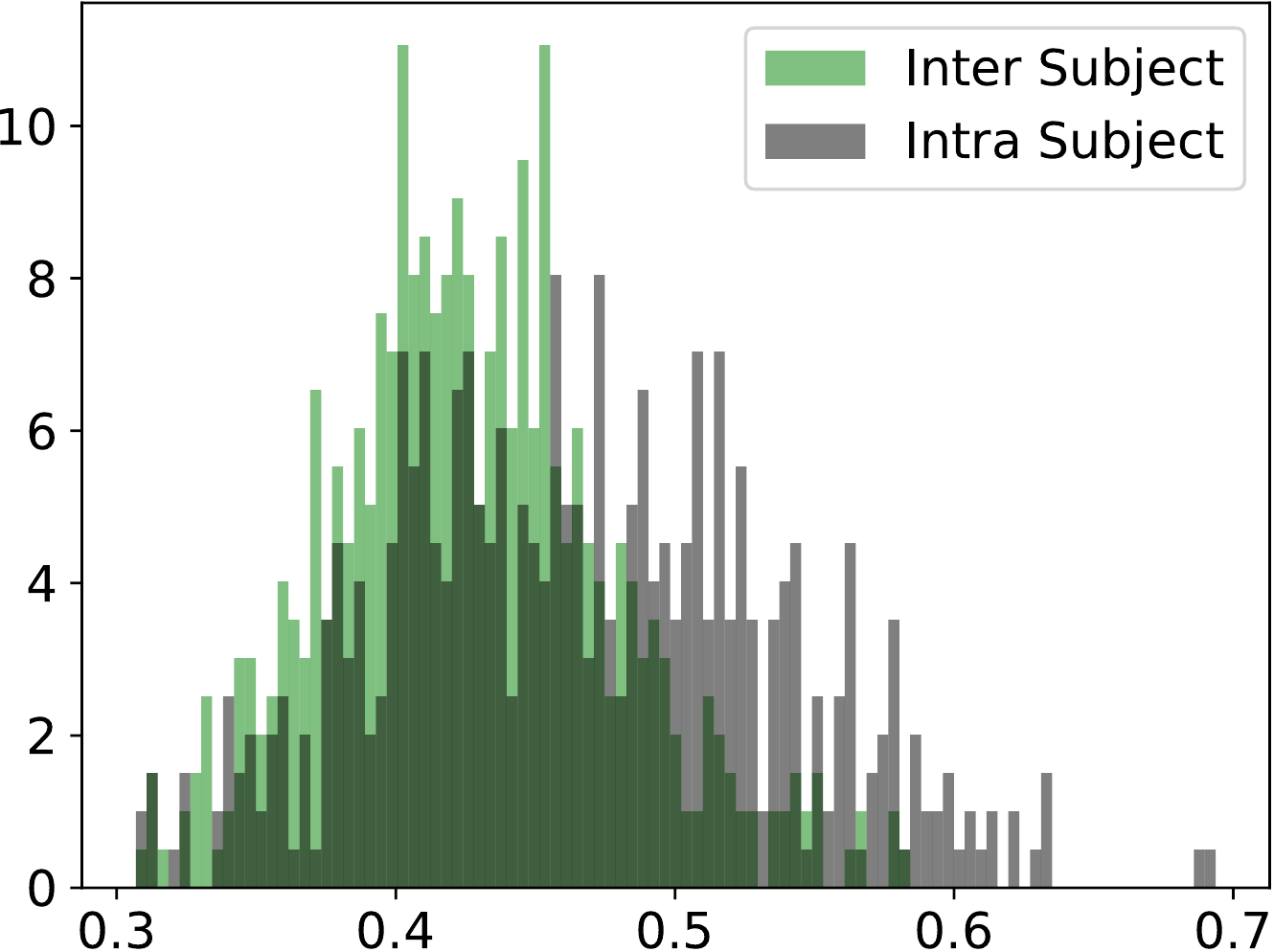} & \includegraphics[width=0.225\linewidth]{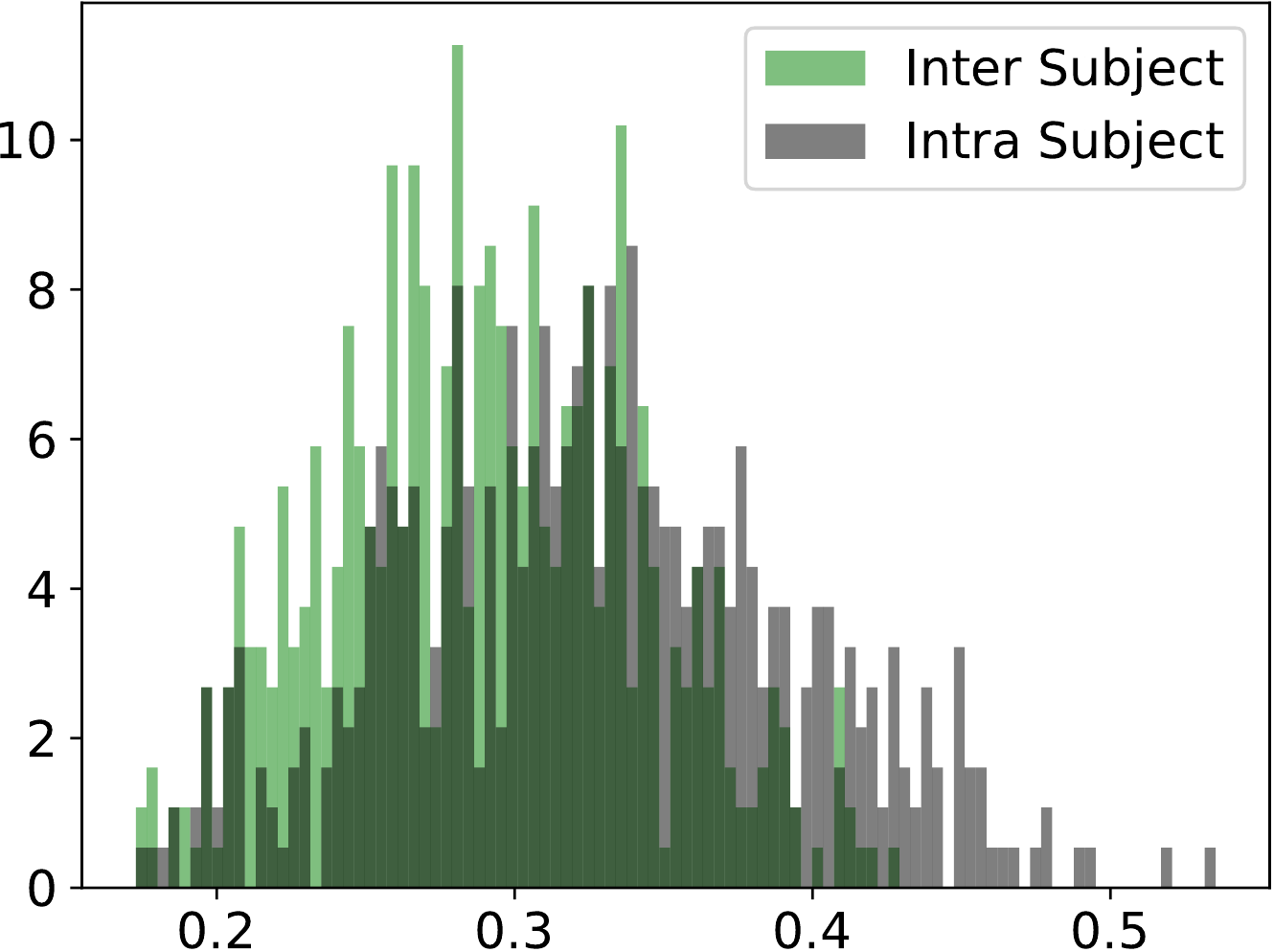} \\
 (a) & (b) & (c) & (d)
\end{tabular}
\end{footnotesize}
\caption{MS-SSIM score and DSC histograms between inter- and intra-subject (a)  undistorted MR images and (b) undistorted segmentation maps (c) deformed images and (d) deformed segmentation maps.}
\label{fig:vis}
\end{center}
\end{figure*} 

\subsection{Evaluation Metrics}
We resort to the Dice similarity coefficient (DSC) to measure the segmentation accuracy of the different methods. Furthermore, to evaluate the methods' ability to obfuscate subject identity, we follow the retrieval-based analysis of~\cite{kim2020privacynet} where an attacker tries to recover the identity of a subject by matching an encoded image or segmentation map against an existing database. In this analysis, mean average precision (mAP) is employed to measure the re-identification accuracy~\cite{Kumar209726}. Let $\xx_i, \yy_i$ be an image and its corresponding segmentation map of a given subject with identity $\mr{id}(i)$, and denote as $\mathcal{T}_{\mr{id}(i)}$ the set of images and segmentation maps of this subject. Also, let $\mathcal{S}^k_i$ be the set of $k$ most similar images to $\xx_i$ or segmentations to $\yy_i$ according to a given re-identification approach. The precision at cut-off $k$ is defined as
\begin{align}
 \big(\mr{precision}@k\big)_i \ = \ \frac{\mathcal{T}_{\mr{id}(i)} \cap \mathcal{S}^k_i}{k}
\end{align}
Considering each proxy test image\,/\,segmentation as a separate retrieval task where one must find other proxy images from the same person, the average precision (AP) for image\,/\,segmentation $i$ is given by
\begin{equation}
 \mr{AP}_i \ = \ \frac{1}{\sum_{j\neq i} s_{ij}} \sum_{k=1}^{N} \big(\mr{precision}@k\big)_i \cdot \widetilde{s}_{ik},
\end{equation}
where $\widetilde{s}_{ik}$ is a label indicating if the $k$-th \emph{most similar} image or segmentation is from the same subject as $i$ or not. mAP is then the mean of AP values computed over all test examples. We also resort to the F1-score to evaluate the performance of the Siamese discriminator.

\subsection{Implementation details}
 
As described in Section \ref{sec:architecture}, we used a U-Net architecture with 3D convolution kernels for the segmentation network and the modified 3D U-Net model of Fig.~\ref{fig:architecture} for the transformation generator. For all experiments, we trained the system for 100 epochs with the Adam optimizer and a learning rate of $10^{-4}$. The balancing weights in Eq. (\ref{eq:total_loss}) were set to $\lambda_1\!=\!0.5$, $\lambda_2\!=\!1$, $\lambda_3\!=\!10$ and $\lambda_4\!=\!1$. The system was implemented in Pytorch, and experiments were performed on Intel(R) Core(TM) i7-6700K 4.0GHz CPU with a 16 GB NVIDIA Tesla P100 GPU. Additional implementation details can be found in the supplementary material. The code will be made public upon acceptance of this paper.


\section{Results}

We start by evaluating the segmentation and re-identification performance of a baseline using non-distorted images of PPMI, which we call \emph{no-proxy baseline}. We then evaluate our privacy-preserving segmentation method on the same data, and conduct an ablation study to measure the contribution of each loss term. Last, we assess our method's ability to generalize on MRBrainS data. 

\subsection{No-proxy baseline}
\label{sec:experiment}

\myparagraph{Re-identification} We measure the ability of the Siamese discriminator 
trained independently to correctly recover the identity of a patient with the original, non-distorted images and segmentation maps of PPMI. These {\em no-proxy} results are reported in the first row of Table~\ref{table:Results}, where we observe that the F1-scores of the discriminator are above $98\%$ and the mAP is close to $100\%$. To further compare the inter\,/\,intra-subject similarity, we computed a MS-SSIM score between each pair of MRI images and each pair of segmentation maps and put the inter\,/\,intra histograms in Fig.~\ref{fig:vis} (a) and (b). As can be seen, when considering non-encoded images, the intra-subject MS-SSIM scores (grey curves) are significantly larger than that of the inter-subjects (green curves). This demonstrates that identity can be recovered easily from non-distorted images.

\begin{figure*}[tp]
\begin{center}
\begin{footnotesize}
\begin{tabular}{cccc}
 
\fig{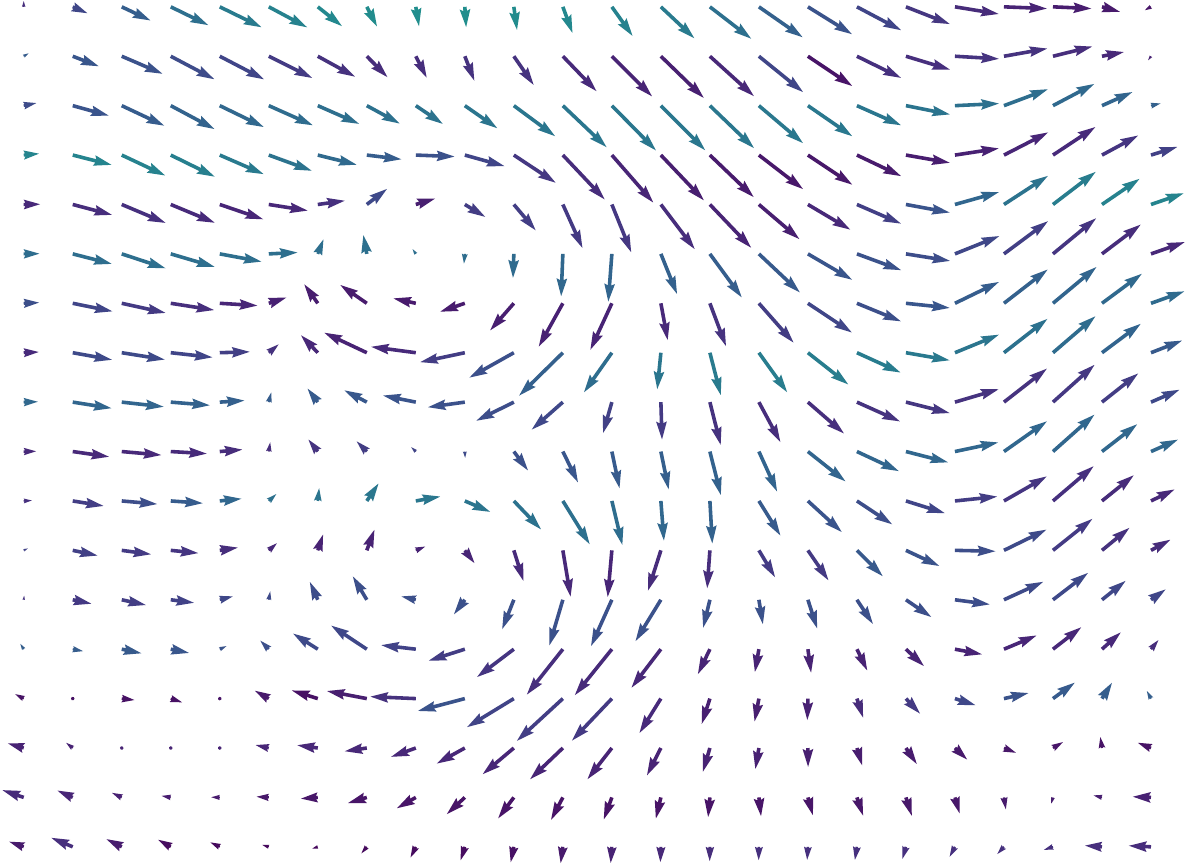} & \fig{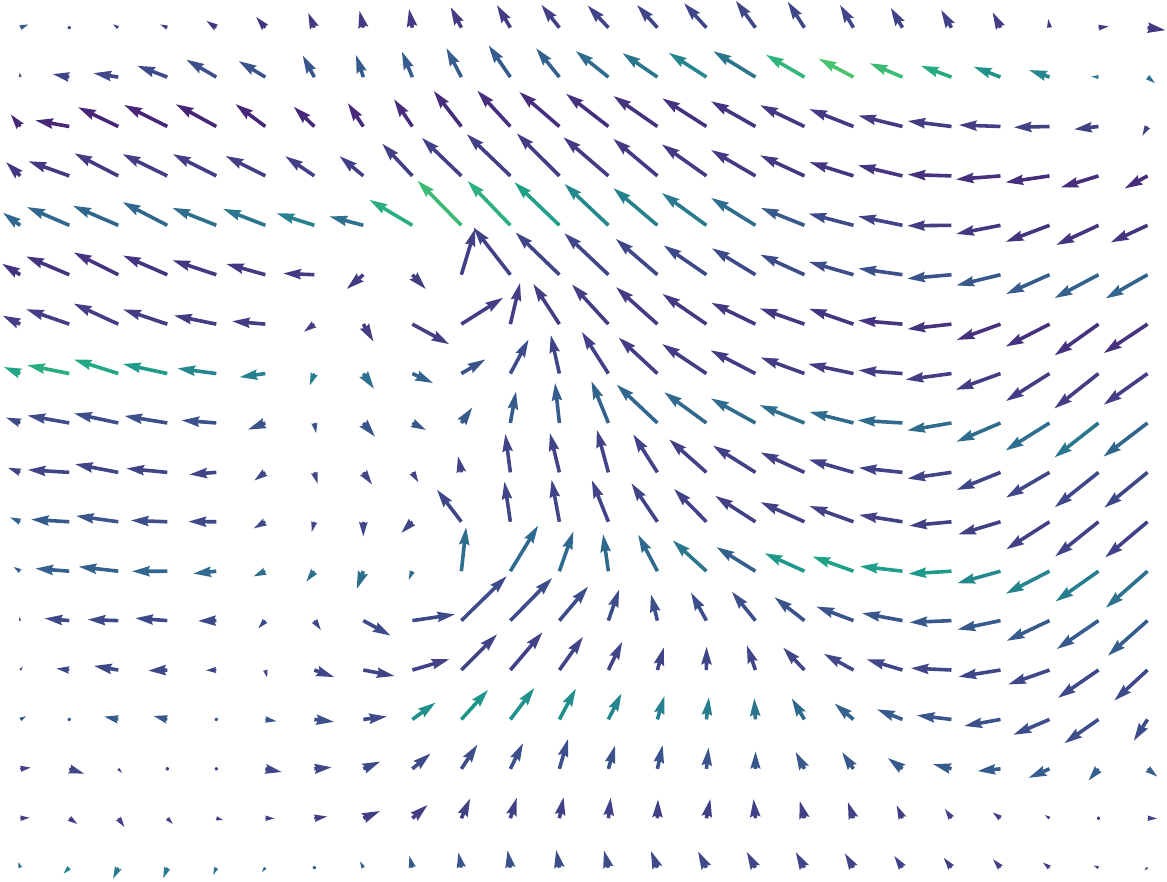} & \fig{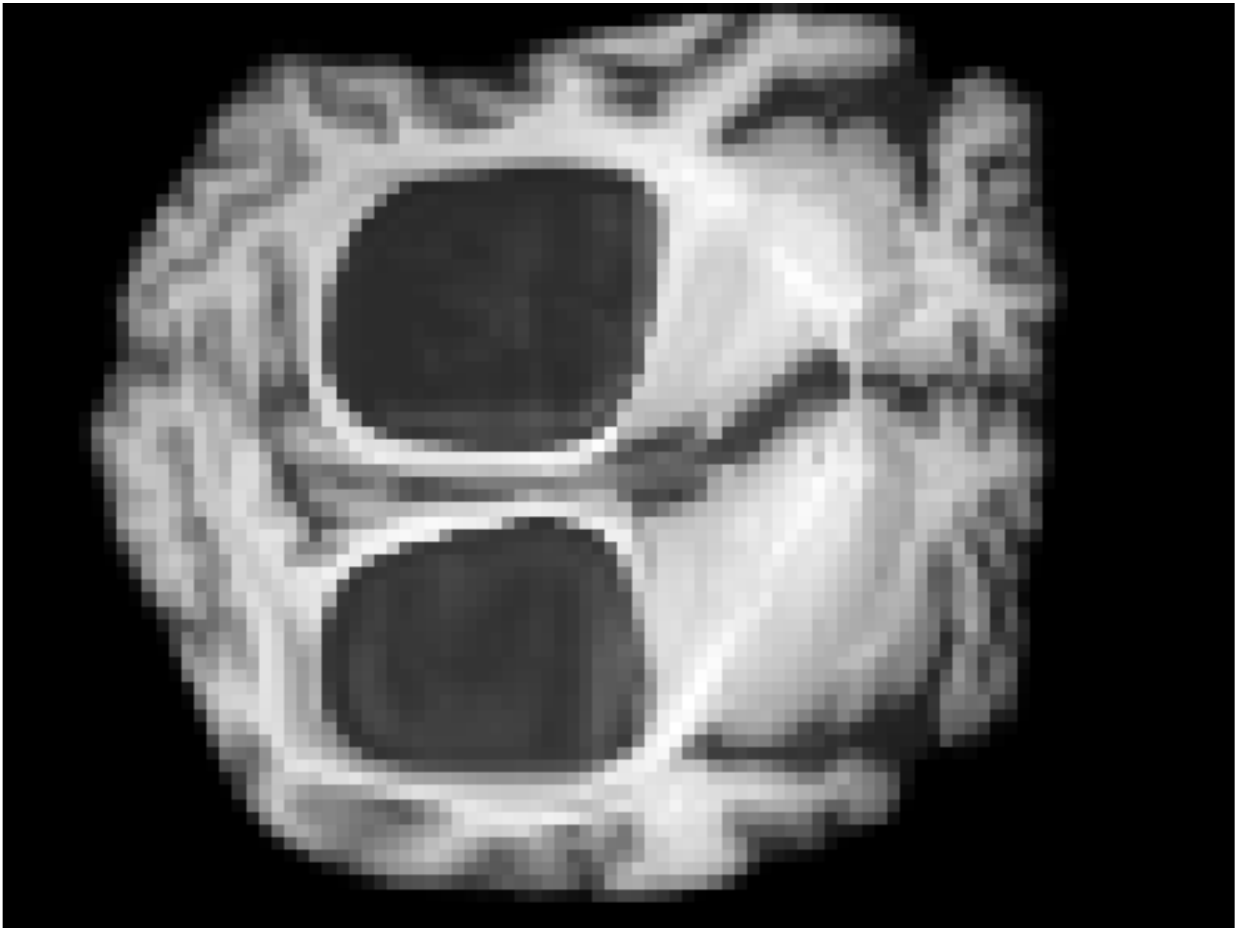} & \fig{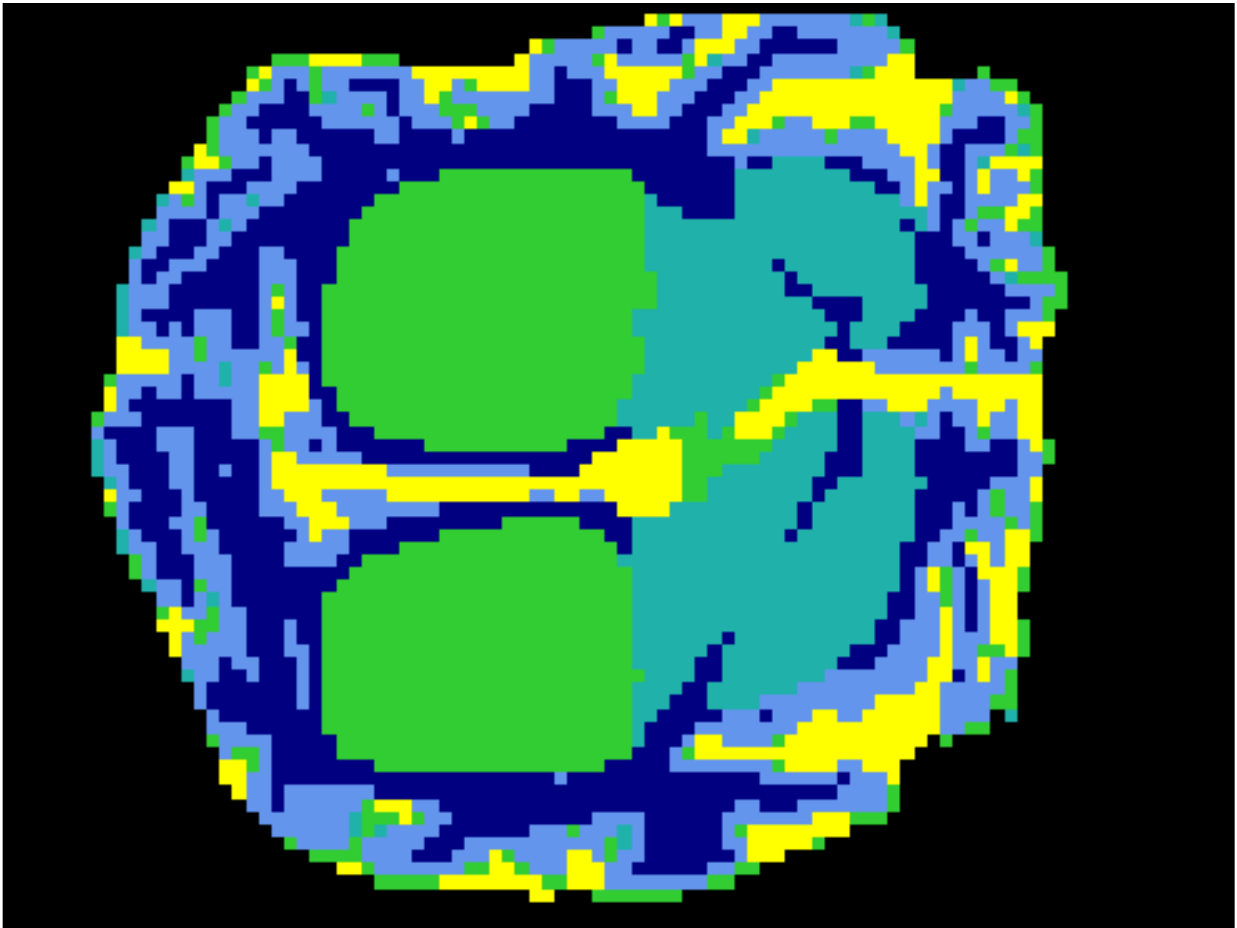} \\
Forward flow-field & Backward flow-field & Deformed image & Deformed segmentation \\[.75em]
 
\fig{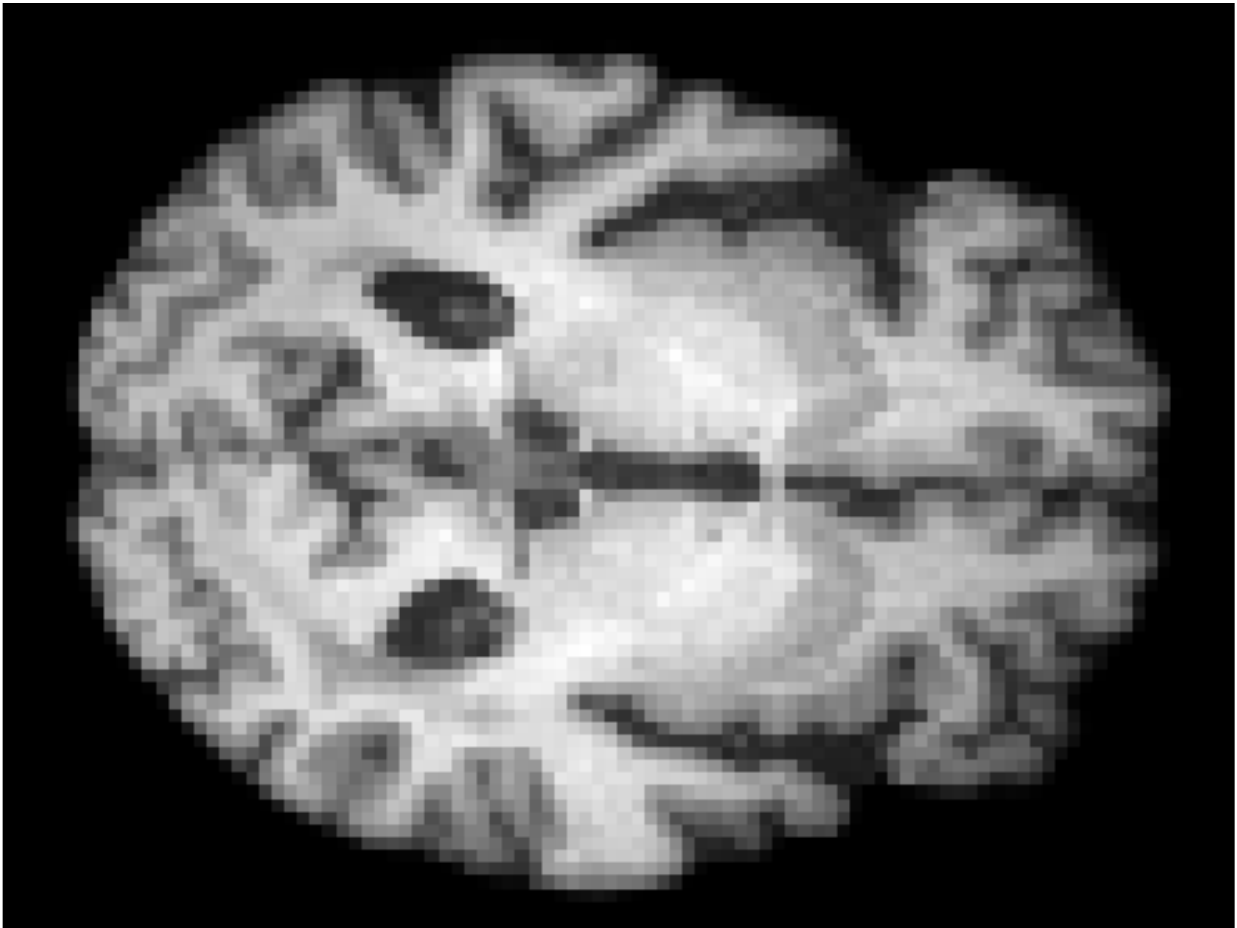} & \fig{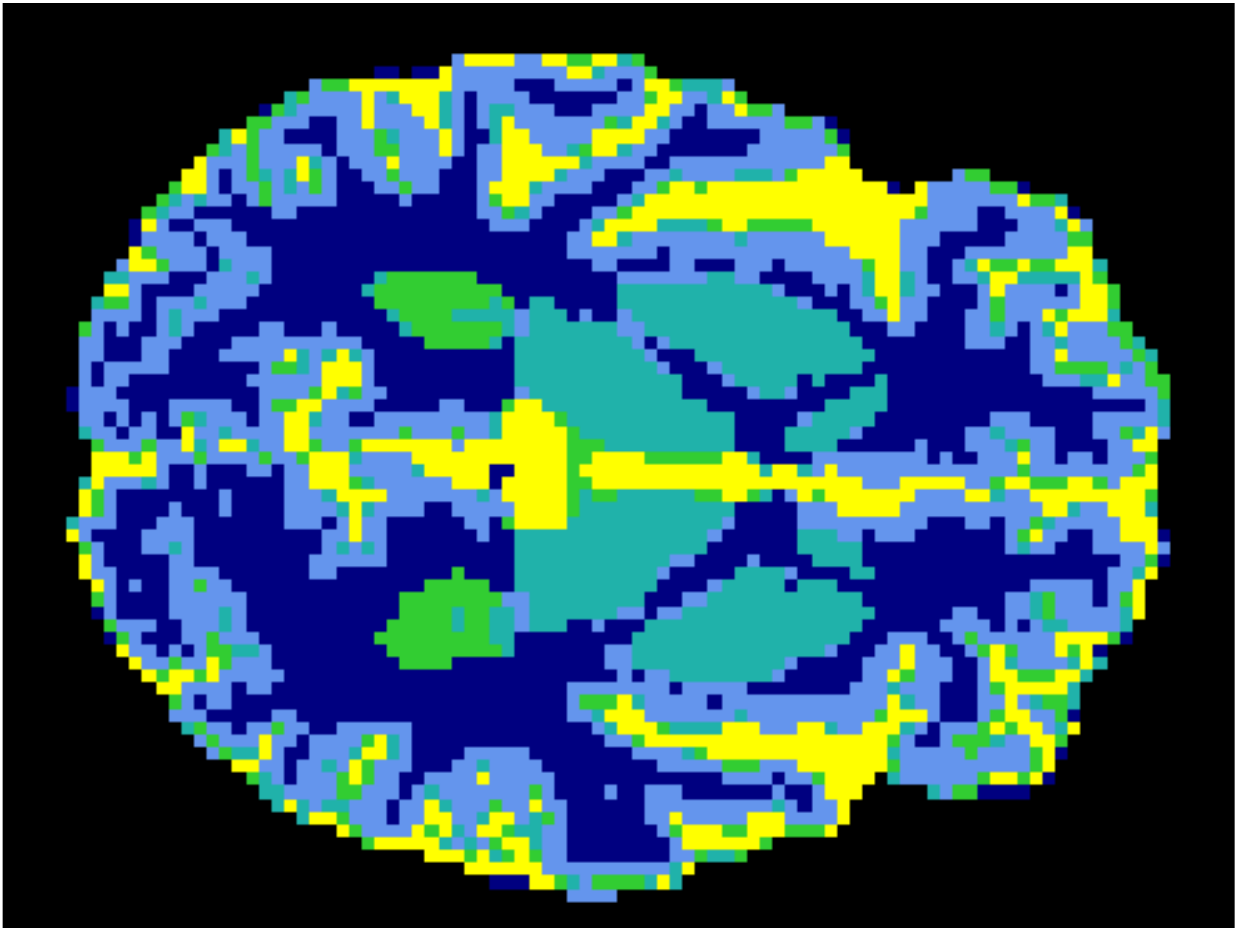} & \fig{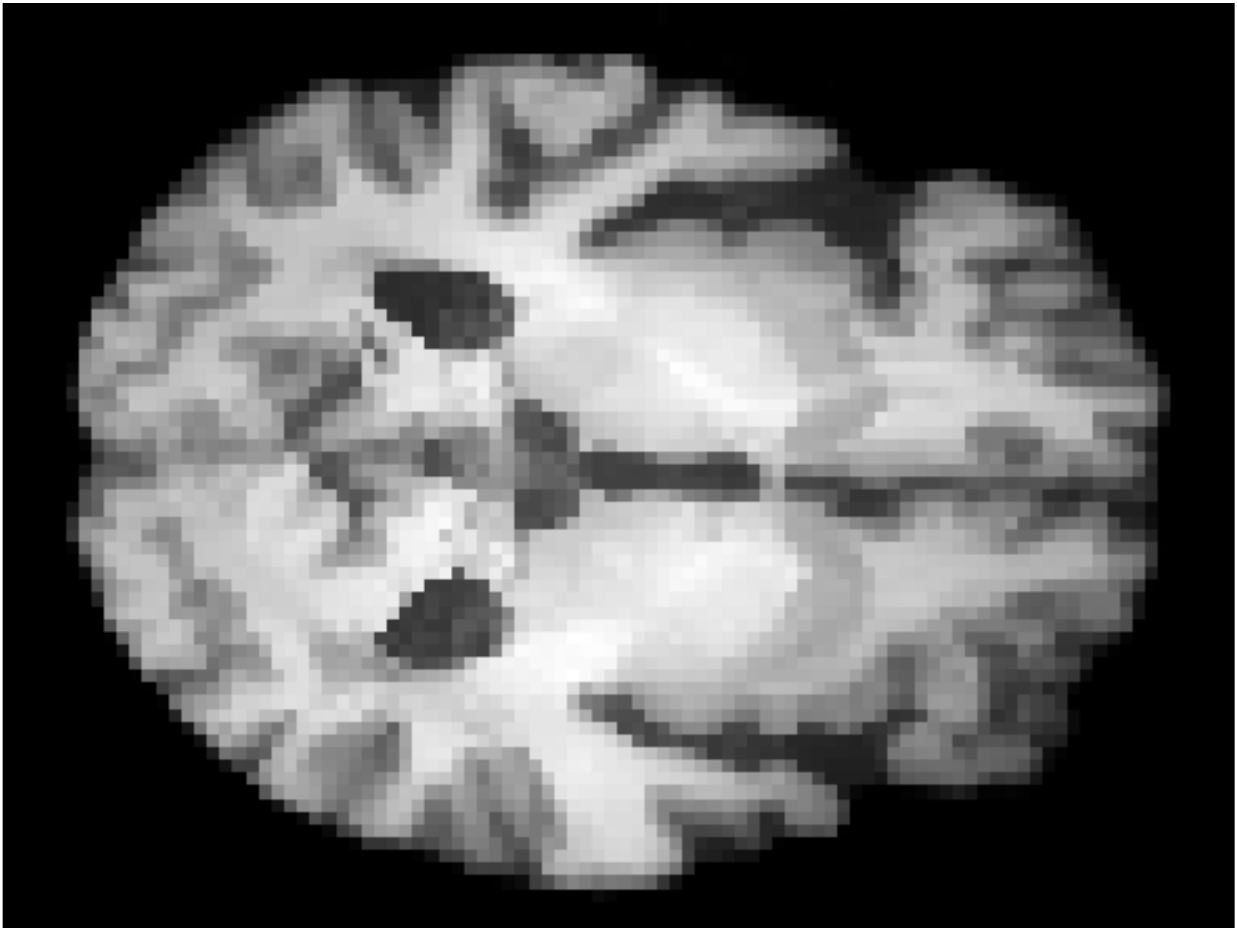} & \fig{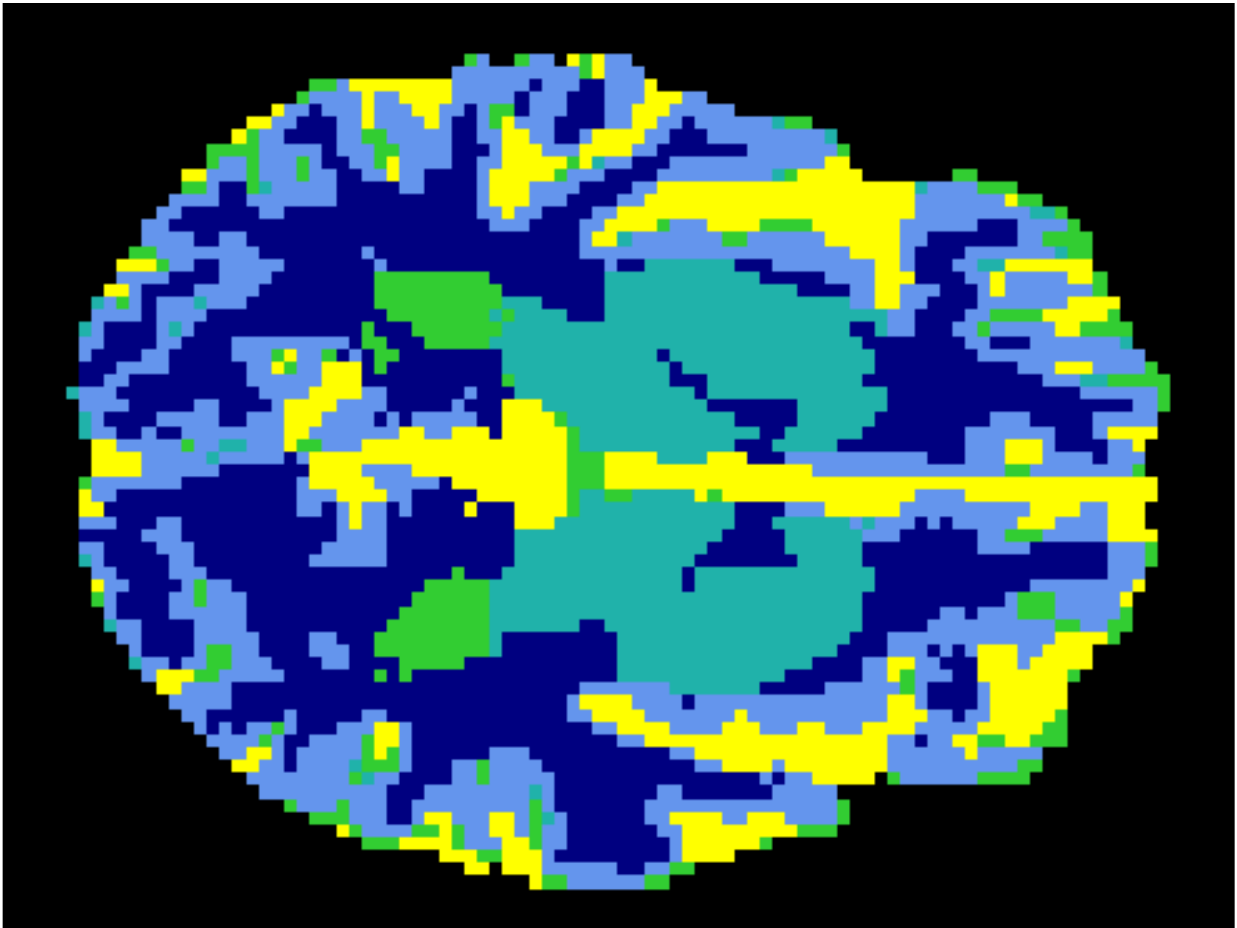} \\
Input image & Segmentation & Reconstructed image & Reconstructed segmentation
\end{tabular}
\end{footnotesize}
\caption{Visualization of forward $f$ and backward $\finv$ deformation fields, input images with its associated ground truth map, deformed image and segmentation map and the reconstructed images and segmentation maps.}
\label{fig:vis2} 
\end{center}
\end{figure*} 

\myparagraph{Segmentation Result} In the {\em no-proxy} row of Table~\ref{table:Results}, we also report the segmentation Dice scores of our segmentation method trained on the undistorted images. The overall Dice score is the average Dice across regions weighted by the region size. These results correspond roughly to those obtained in a recent publication for a similar architecture~\cite{dolz2019}. Note that the nuclei and the internal CSF have a lower Dice due to their smaller sizes.

\subsection{Results on PPMI}

\myparagraph{Re-identification} 
Here, we measure the ability of the generator to obfuscate the identity of a patient. Quantitative results for our system are reported on the {\em All losses} row of Table~\ref{table:Results}.
We can see that the F1-scores drop to $5\%$ and the mAP to $9\%$ for both the distorted image and distorted segmentation maps. This indicates that most information on patient identity has been removed from these data. Fig.~\ref{fig:vis} (c) and (d) gives the inter\,/\,intra-subject MS-SSIM score histograms between of deformed images and deformed segmentation map. We observe that the grey and green curves overlap almost entirely, showing that same-subject images are as different as those from separate subjects.  Figure~\ref{fig:vis2} depicts an input image and segmentation ground-truth, together with their associated flow-fields, distorted and reconstructed images. Despite the large and variable deformation applied to images, the segmentation network can precisely delineate the complex-shaped brain regions.

\myparagraph{Reconstruction}
The first row of Table~\ref{table:Reconstruction_Results}, i.e., \textit{All losses}, reports the reconstruction accuracy obtained with both input images and segmentation maps. MS-SSIM is used to evaluate the similarities on the raw inputs, whereas we employ the Dice score to measure differences on the segmentation maps. Particularly, we observe that our system is capable of reconstructing both distorted images and segmentation masks, with a MS-SSIM value near to 100\% and an overall Dice above 0.98. 


\myparagraph{Segmentation}
The segmentation DSC achieved by our method is reported in the {\em All losses} row of Table~\ref{table:Results}. We want to highlight that even though the segmentation network in the proposed system underperforms the network trained on undistorted images, a DSC higher than 0.80 is suitable for several clinical applications. This is supported by observations in the clinical literature, where some authors report DSC values of 0.70 to be acceptable~\cite{363096,article1,pmid23336255,article2} while others, more conservative, suggest minimum DSC values of 0.80~\cite{Mattiucci2013AutomaticDF}.  

\myparagraph{Comparison to the state-of-art} We also compared our system to the recently-proposed Privacy-Net~\cite{kim2020privacynet}. As can be seen, while the segmentation Dice scores are globally similar to those of our approach ({\em all loss} row), our re-identification mAP values are significantly lower both on images and segmentation maps. Note that both the system in \cite{kim2020privacynet} and the proposed framework resort to UNet as backbone segmentation architecture. This demonstrates that \textit{i)} our approach preserves the segmentation capabilities shown in \cite{kim2020privacynet}, and also \textit{ii)} it can drastically improve the obfuscation of identity.


\subsection{Ablation study}

To examine the importance of each loss term, we proceeded to the following ablation study.

\begin{figure}[]
\begin{center}
\begin{footnotesize}
\begin{tabular}{cc}
\includegraphics[width=.20\textwidth]{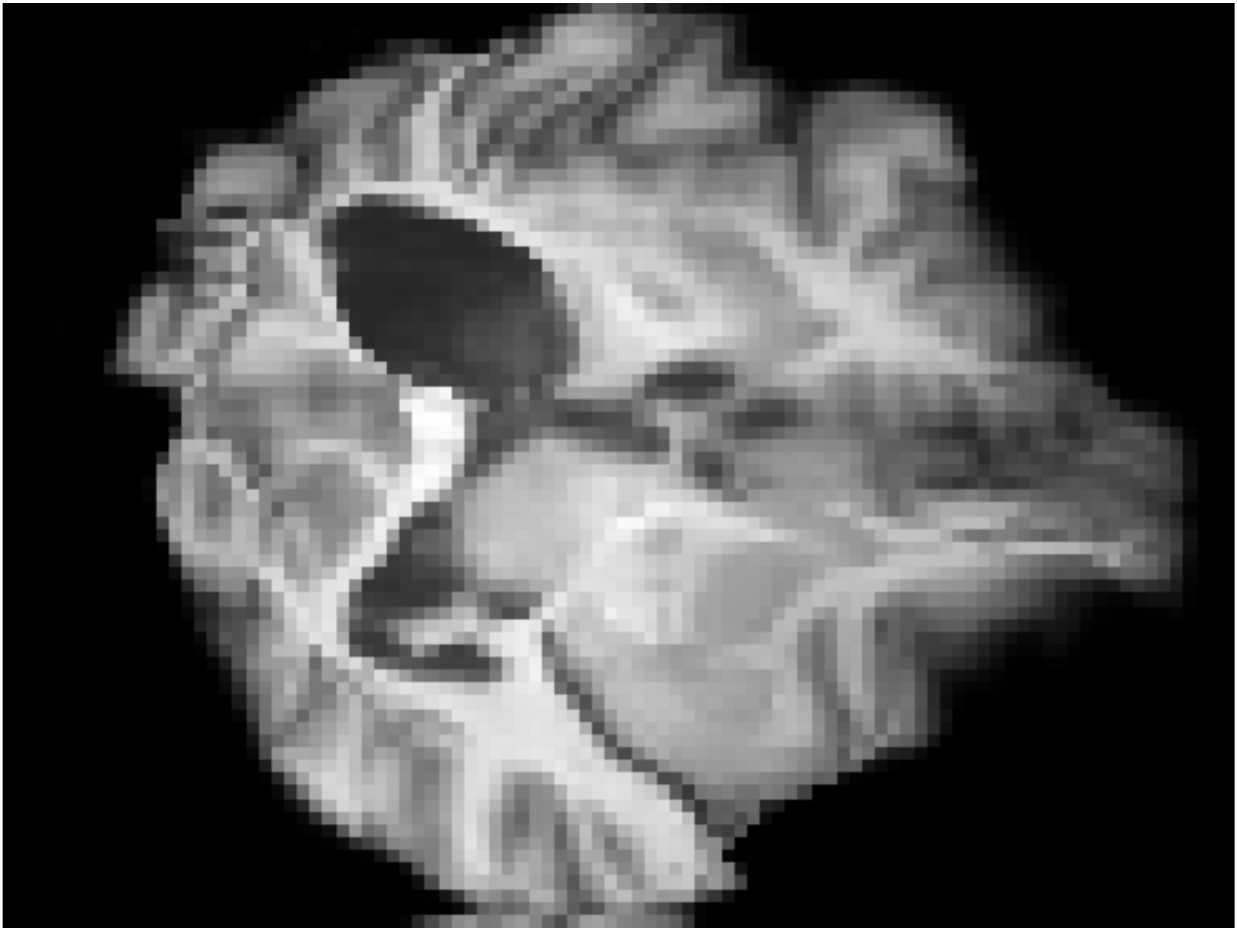} & \includegraphics[width=.20\textwidth]{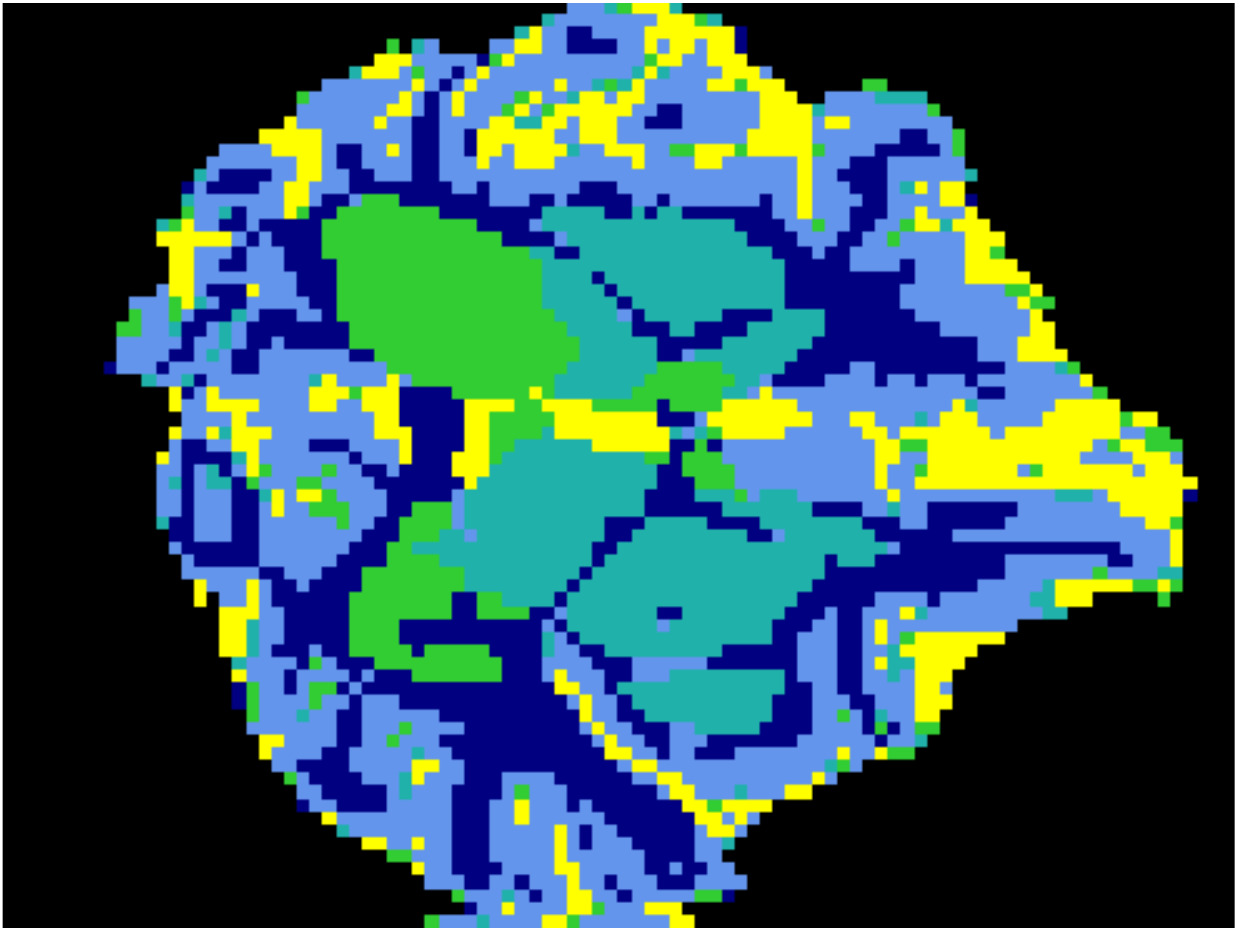} \\
Deformed image & Deformed segmentation map \\
\includegraphics[width=.20\textwidth]{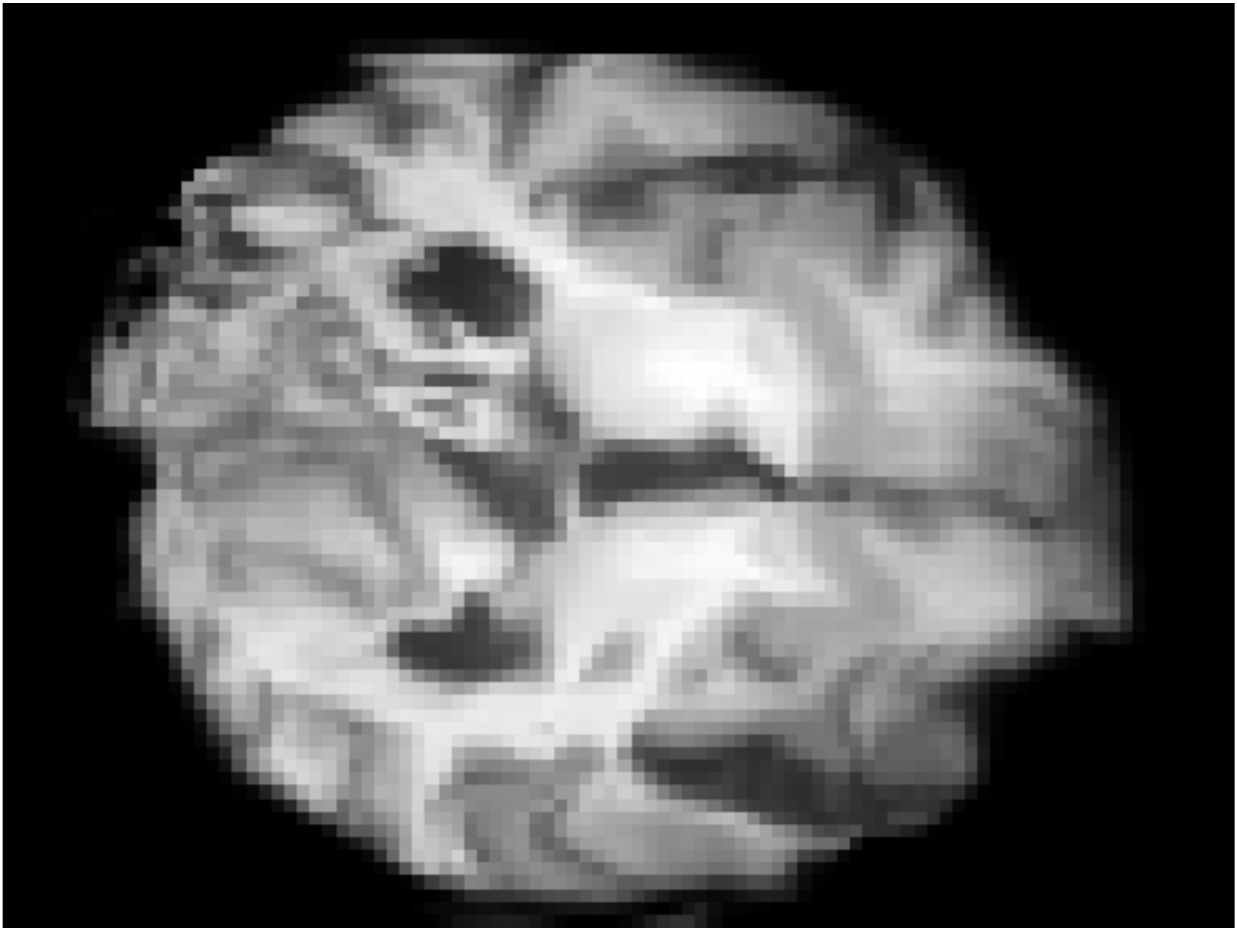} & \includegraphics[width=.20\textwidth]{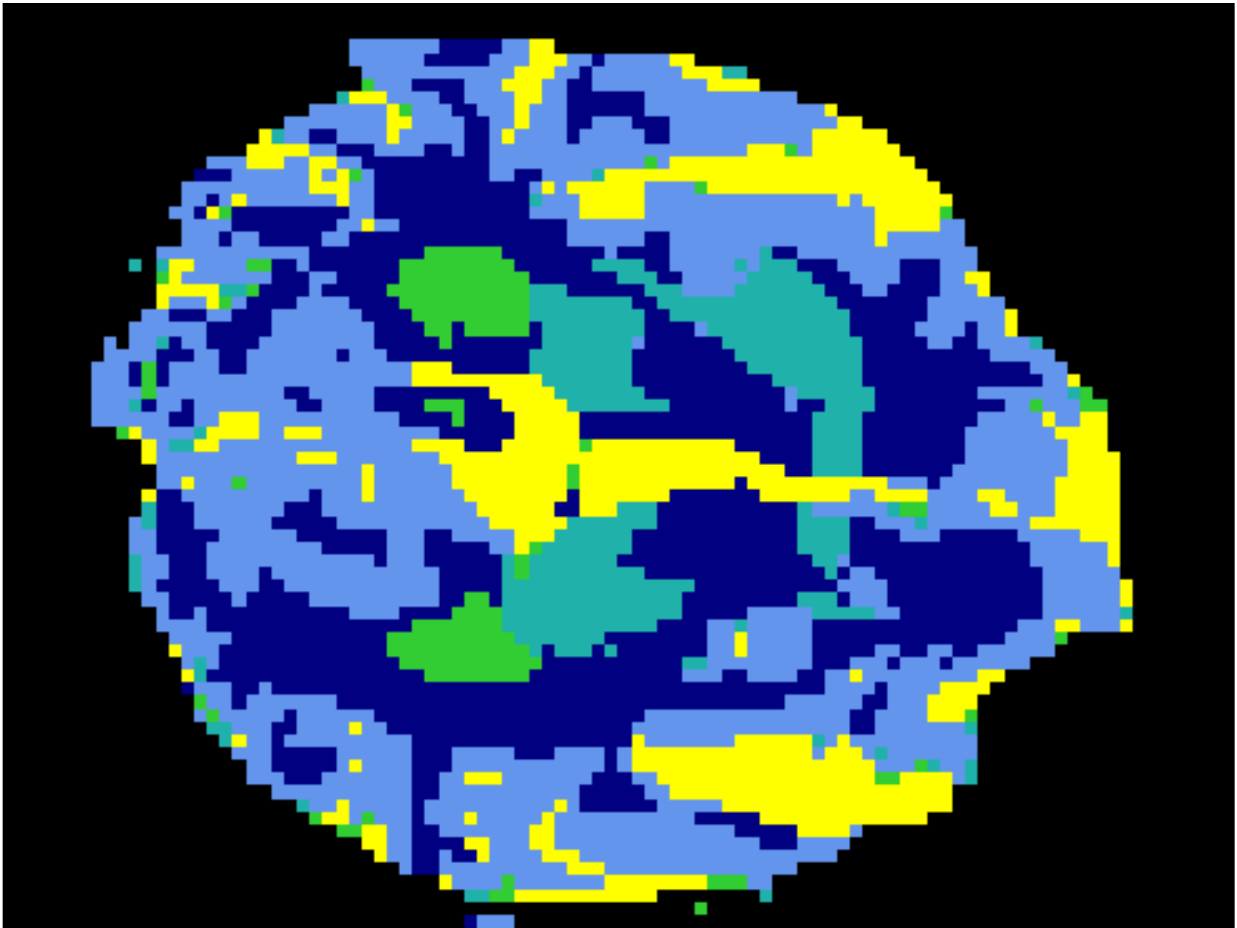} \\
 Reconstructed image & Reconstructed segmentation map
\end{tabular}
\end{footnotesize}
\caption{Reconstructed image and segmentation map of a deformed brain without the invertibility loss of Eq.~(\ref{eq:invert_loss}).}
\label{fig:vis_wo_inv}
\end{center}
\end{figure}

\myparagraph{Invertibility loss} We trained the whole system without the invertibility loss  of Eq.~(\ref{eq:invert_loss}). Although the segmentation loss in Eq.~(\ref{eq:segmentation_loss}) implicitly handles the reconstruction of segmentation maps, it is not sufficient for learning a reversible transformation. As can be seen from  Tables~\ref{table:Results} and \ref{table:Reconstruction_Results}, the reconstruction accuracy  and the segmentation Dice score for this setting are catastrophically low. This is further illustrated in Fig.~\ref{fig:vis_wo_inv} were the reconstructed image and segmentation map of a deformed brain are plagued with artifacts.

\begin{figure}[]
\begin{center}
\begin{footnotesize}
\begin{tabular}{cc}
\includegraphics[width=.20\textwidth]{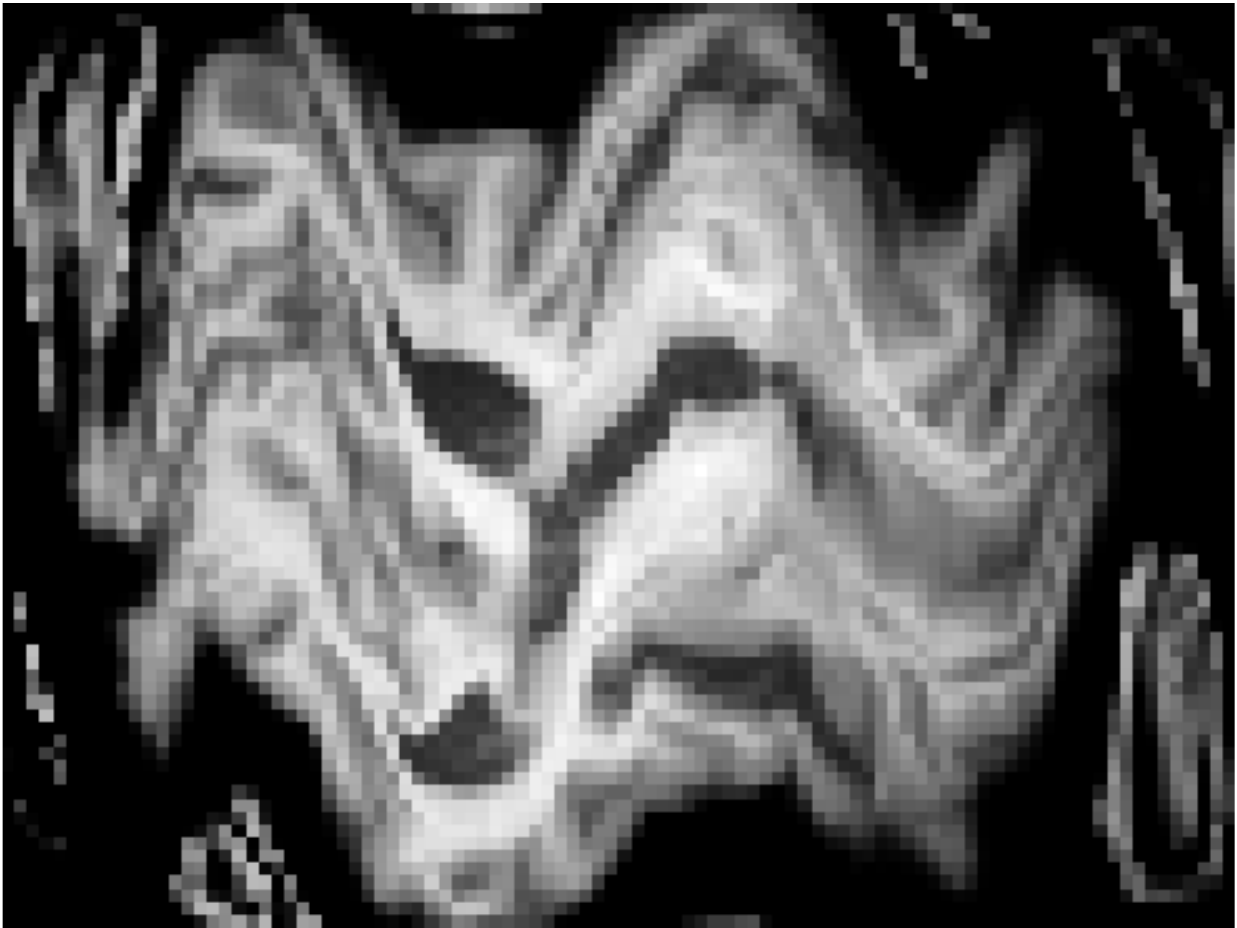} & \includegraphics[width=.20\textwidth]{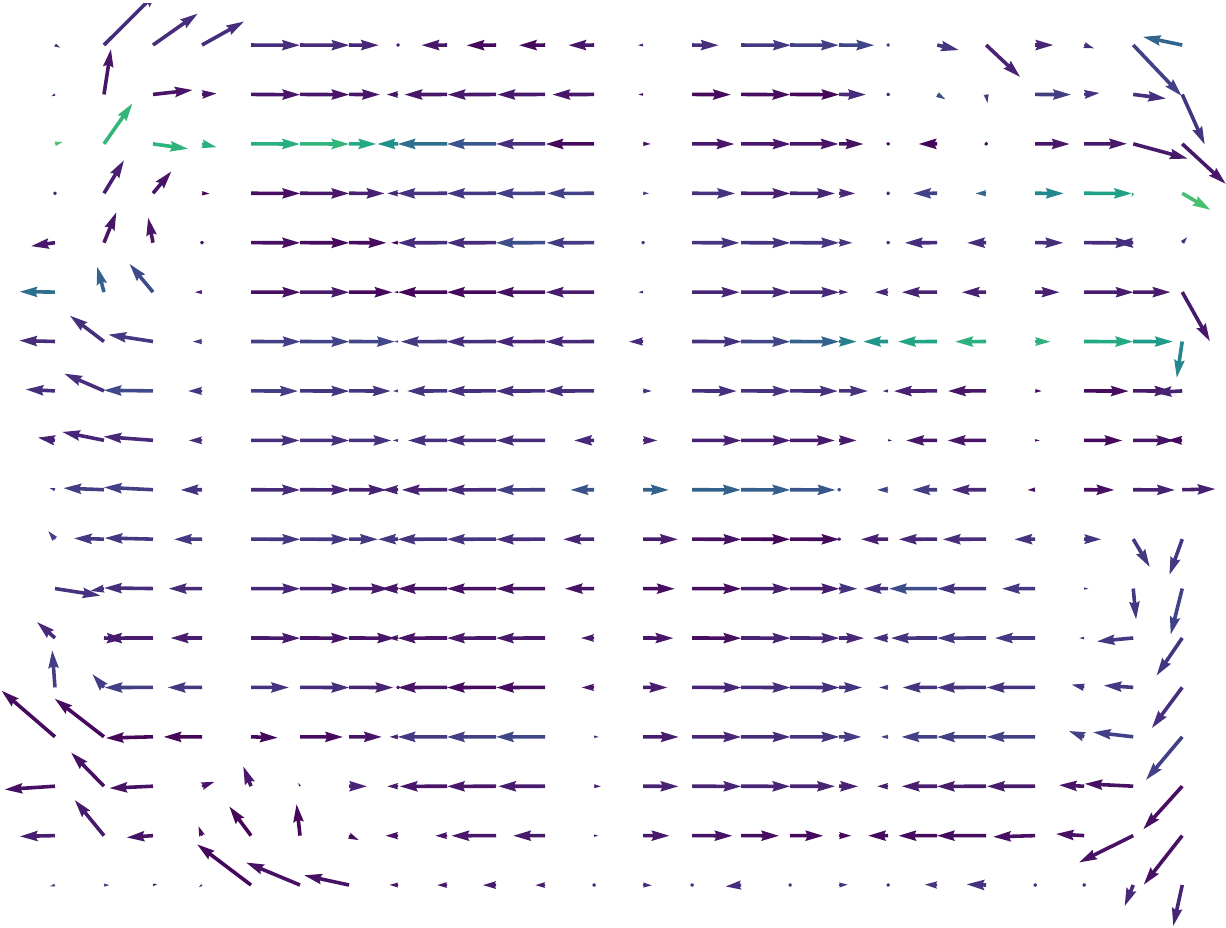} \\
 Artifact deformed image & Discontinuity flow-field
\end{tabular}
\end{footnotesize}
\caption{Artifact in the deformed image when train without smoothness.}
\label{fig:vis_wo_smooth}
\end{center}
\end{figure}

\myparagraph{Smoothness loss} We trained the system without the smoothness loss of Eq.~(\ref{eq:smoothness}) that regularizes the flow-field. As shown in Fig.~\ref{fig:vis_wo_smooth}, the resulting flow-field has abrupt discontinuities which degrade the reconstruction accuracy and lead to a drop in accuracy as reported in Tables~\ref{table:Results} and \ref{table:Reconstruction_Results}.

\begin{figure}[]
\begin{center}
\begin{footnotesize}
\begin{tabular}{cc}
\includegraphics[width=.20\textwidth]{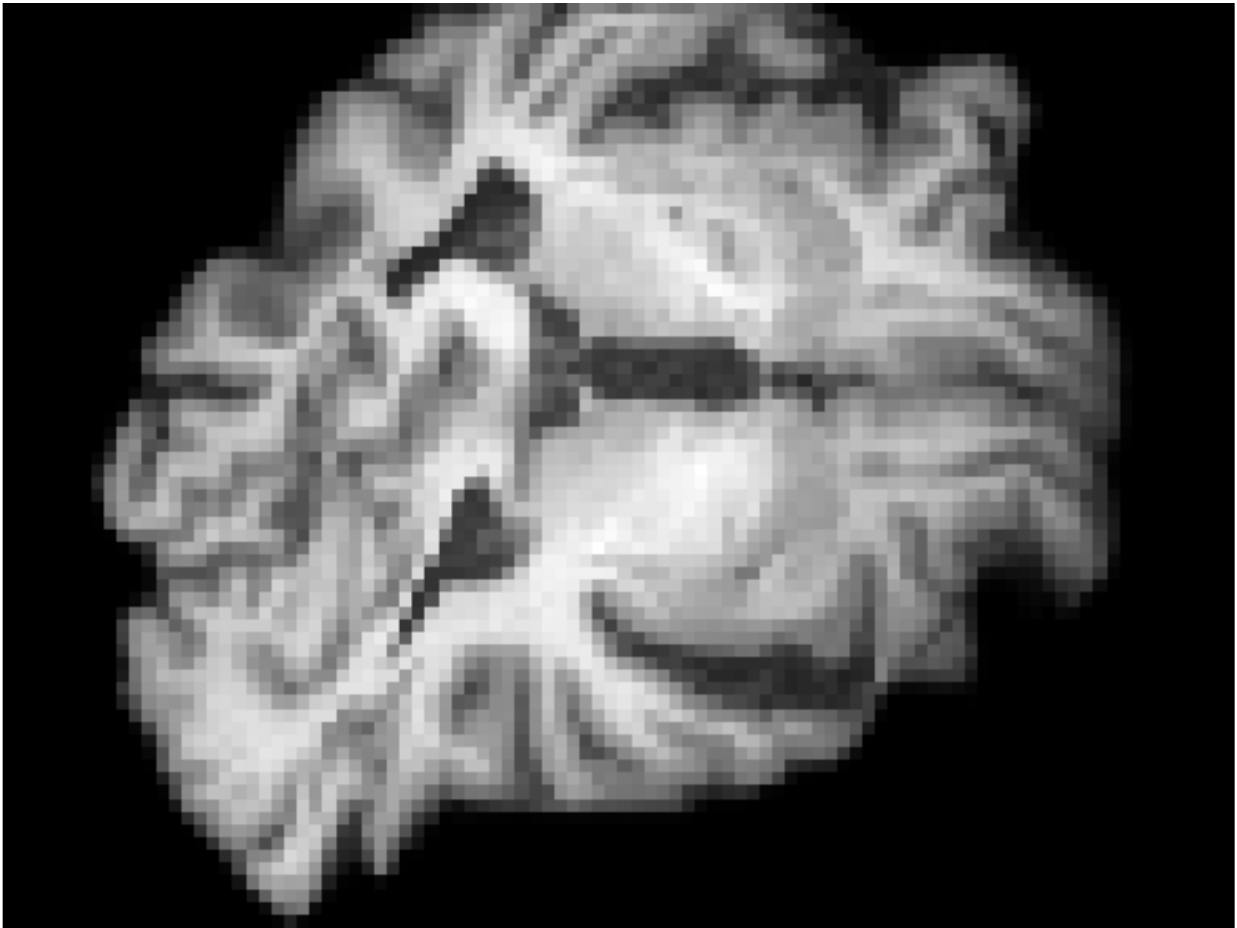} & \includegraphics[width=.20\textwidth]{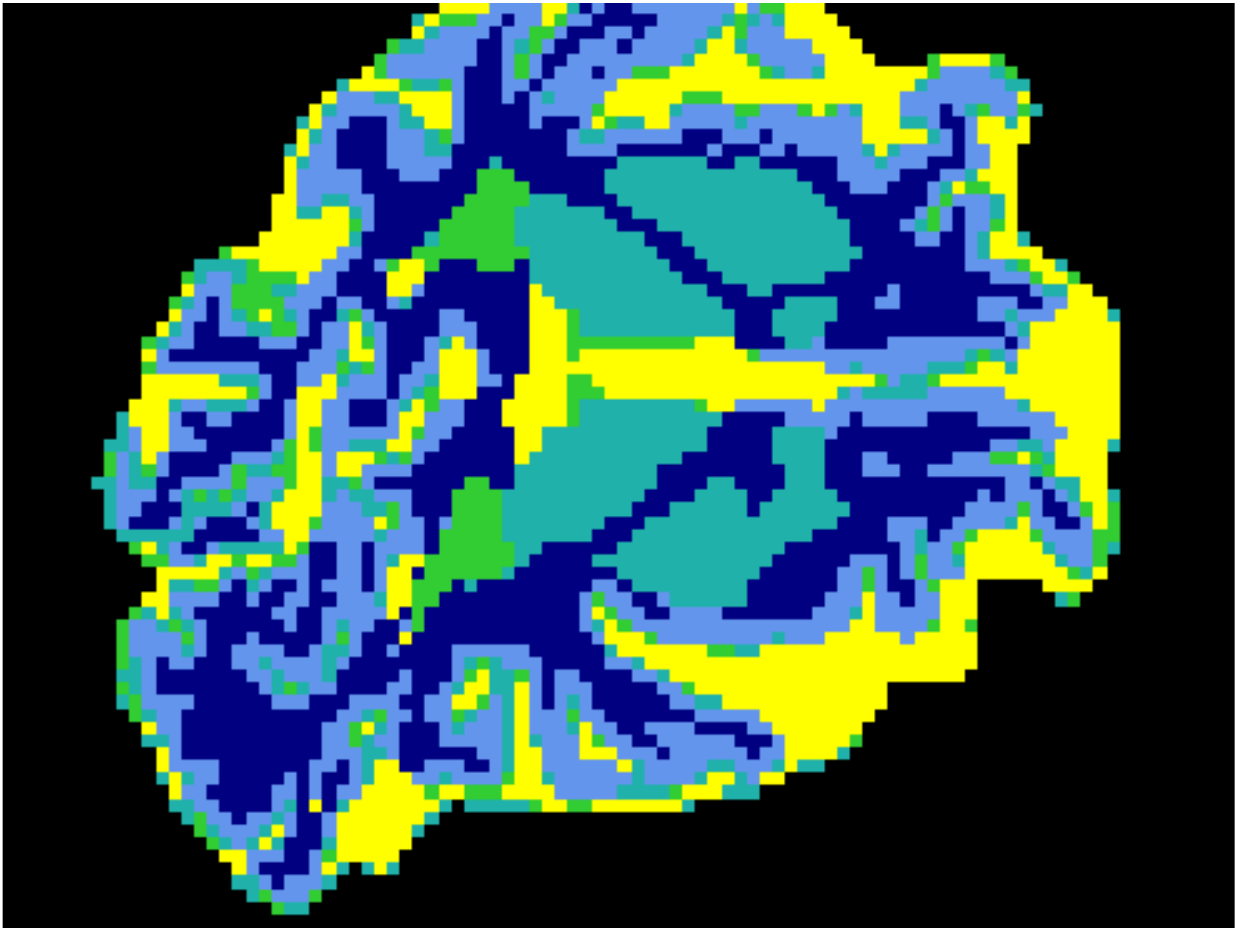} \\
 Deformed image & Deformed segmentation map
\end{tabular}
\end{footnotesize}
\caption{Weak distortion when train without diversity loss}
\label{fig:vis_wo_diversity}
\end{center}
\end{figure}

\myparagraph{Diversity loss} As indicated in Tables~\ref{table:Results} and \ref{table:Reconstruction_Results} and illustrated in Fig.  ~\ref{fig:vis_wo_diversity}, removing the transformation diversity loss of Eq. ~(\ref{eq:distorion_loss}) leads to a higher reconstruction accuracy and Dice score. While this might seem beneficial, it comes at the expense of a much higher re-identification F1-score and mAP as shown in the last row of the Table~\ref{table:Results}.


\subsection{Results on MRBrainS}


To demonstrate the generalizability of the learned transformation for privacy-preserving segmentation, we fixed the generator pre-trained on PPMI and then only retrained the segmentation network on the MRBrainS data. Table~\ref{table:MRBrainS_Results} reports the segmentation accuracy for non-distorted and distorted  images of MRBrainS. Similarly to PPMI, we also observe a small drop of the Dice score between the segmentation results without and with deformation. Particularly, our method achieves an overall Dice of $83.9\%$, which is nearly 4\% lower than the performance on non-deformed images. This suggests that the proposed approach can generalize well to other datasets.

\begin{table}[]
\centering
\caption{Segmentation result on the MRBrainS test set.}
\label{table:MRBrainS_Results}
\small
\begin{tabular}{lcccc}
\toprule
Setting & Overall & GM & WM & CSF \\
\midrule
Non-distorted images & 0.881 & 0.879 & 0.887 & 0.883 \\
Distorted images & 0.839 & 0.832 &0.840 & 0.835 \\
\bottomrule
\end{tabular}
\end{table}



\section{Conclusion}

We presented a strategy for learning image transformation functions that remove sensitive patient information from medical imaging data, while also providing competitive results on specific utility tasks. Particularly, our system integrates a flow-field generator that produces pseudo-random deformations on the input images, removing structural information that otherwise could be used to recover the patient identity from segmentation masks. This contrasts with prior works, where the image deformations come in the form of intensity changes, leading to the preservation of identifiable structures. This was empirically demonstrated in our experiments, where the proposed system drastically decreased the re-identification performance based on segmentation masks, compared to competing methods. Additional numerical experiments suggest that the proposed approach is a promising strategy to prevent leakage of sensitive information in medical imaging data.

\bibliographystyle{unsrt}
\bibliography{egbib}

\end{document}